\documentstyle[aps,preprint,epsfig,tighten]{revtex}
\begin{document}
\thispagestyle{empty}
\noindent\hspace*{\fill}  hep-ph/yymmdd \\
\noindent\hspace*{\fill}  \today   \\

\begin{center}\begin{Large}\begin{bf}
Self-dual homogeneous gluon field and electromagnetic structure of pion.   \\

\end{bf}\end{Large}\vspace{.75cm}

 \vspace{0.5cm}
{\bf Ja. V. Burdanov}, \\
Laboratory of Nuclear Problems, \\
Joint Institute for Nuclear Research, 141980 Dubna,
Russia, \\

{\bf G. V. Efimov}, \\
Bogoliubov Laboratory of Theoretical Physics, \\
Joint Institute for Nuclear Research, 141980 Dubna,
Russia, \\

and {\bf S. N. Nedelko}, \\
Bogoliubov Laboratory of Theoretical Physics, \\
Joint Institute for Nuclear Research, 141980 Dubna,
Russia \\

\end{center}
\vspace{1cm}\baselineskip=35pt

\begin{abstract}
        The transition form factor $F_{\gamma\pi}(Q^2)$,
         decay width $\Gamma(\pi^0\to\gamma\gamma)$ and
         charge form factor of pion are calculated
        within the model of induced nonlocal quark currents
based on the assumption that the nonperturbative QCD vacuum can be
characterized by a homogeneous (anti-)self-dual gluon field.
        It is shown that the interaction of the quark spin with the
         vacuum  gluon field,
        being responsible for the chiral symmetry breaking and
        the spectrum of light mesons,
          can also play the decisive role
         in forming the form factor $F_{\gamma\pi}(Q^2)$ and
         decay width $\Gamma(\pi^0\to\gamma\gamma)$.
        An asymptotic behavior of quark loops in the presence of the
        background gluon field for large  $Q^2$ is discussed.

\end{abstract}
\newpage\baselineskip=18pt

\section{Introduction}

Pion electromagnetic properties were intensively studied
within different theoretical approaches, such as perturbative QCD~\cite{br},
the factorization theorems and QCD sum rules~\cite{radyush}-\cite{ioffe},
approaches based on the Bethe-Salpeter
equation~\cite{roberts}-\cite{kissl}, the instanton liquid
model \cite{dor},
relativistic quark
models~\cite{gross,valera},
nonlocal quark model with confinement~\cite{efiv},
and some others.
The reason for this permanent interest is quite understandable:
a variety of general features of strong interactions
clearly manifest themselves in pion physics. First of all this
relates to the mechanisms of chiral symmetry breaking in QCD
and general behaviour of electromagnetic form factors of
hadrons. The charged form factor
played the particularly important role for explanation
of the quark counting rules~\cite{matv} and factorization
hypothesis within QCD
(for review see, e.g.~\cite{radyush}). The two-photon decay of neutral pion
and the transition form factor relates to the triangle anomaly
and is important for testing the models
of QCD vacuum~\cite{br,rad1}.

The present paper is devoted to description of
electromagnetic properties of pions within
the model of induced
nonlocal quark currents developed in our previous papers~\cite{efned,bur}.
Namely, we calculate the charge and transition form factors and two-photon
decay constant of pion.
The model is based on the  assumption that
the (anti-)self-dual homogeneous field~\cite{leutw,mink,eliz,fin}
\begin{eqnarray}
\label{field}
\hat B_\mu(x)=\hat n B_{\mu\nu}x_\nu, \
\hat n=\lambda_3\cos\xi + \lambda_8\sin\xi,
\nonumber\\
\tilde B_{\mu\nu}=\pm B_{\mu\nu}, \
B_{\mu\rho}B_{\rho\mu}=-B^2\delta_{\mu\nu}
\end{eqnarray}
can be considered as a dominating gluon configuration in the QCD vacuum.
In other words, it is assumed that the effective potential for QCD
has a minimum at nonzero strength of the background field~\cite{mink,eur}.

Vacuum field (\ref{field}) leads to spontaneous violation
of a range of symmetries such as CP, colour and O(3).
A satisfactory restoration of these symmetries at
the hadronic scale assumes an existence of
domain structures in the vacuum. In
a given domain the vacuum field has a specific direction
and is either self-dual or anti-self-dual, but
this is uncorrelated with the specific realization
of Eq.(\ref{field}) in another domain.
The idea of domains in the QCD vacuum was discussed in
application to various homogeneous fields~\cite{leutw,Amb80,AmS90,chak}.
In complete theory the domain walls
should be describable by some appropriate
classical solitonic solutions of equations of motion.
In the effective model under consideration
this idea is realized as a prescription that
different quark loops (namely, those separated by the meson lines)
in a diagram must be averaged over different configurations of the vacuum
field~(\ref{field}) independently of each other.

A consideration of both the
quark-gluon dynamics in the external field (\ref{field})
and its manifestations in meson properties indicates that
this field provides for a picture of confinement
and chiral symmetry breaking which is unexpected but
looks quite self-consistent and agrees with meson phenomenology.

Being taken into account nonperturbatively,
the vacuum field under consideration changes analytical
properties of the quark and gluon propagators drastically
and makes QCD to be a nonlocal
quantum field theory.
The propagators of color charged
fields in the presence of field (\ref{field}) are entire
analytical functions in the complex momentum
plane~\cite{efiv,efned,leutw,eur}
which means an absence of quarks and gluons in the observable spectrum
of hadrons. The propagators are modified essentially in the infrared
region and show the usual
behaviour at short distances.
Consideration of the
Wilson loop in the presence of this field shows that the
Wilson criterion is satisfied~\cite{efnedcal} with an oscillator potential
between heavy quarks. The potential arises effectively due to an interaction
of the charges with vacuum field but not by virtue of gluon exchange.

Due to the (anti-)self-duality and homogeneity of the field
(\ref{field}) the operator $\gamma_\mu(\partial_\mu-iB_\mu)$
has an infinite set of zero eigenmodes
whose contribution to the quark propagator
leads to the relation between densities of quark
and gluon condensates:
\begin{eqnarray}
\label{gluequa}
\sum\limits_{f=1}^{N_{\rm F}}
\lim_{m_f\to0}m_f\langle\bar q_f(x)q_f(x)\rangle=
-N_{\rm F}\frac{B^2}{2\pi^2},
\end{eqnarray}
which is valid to all loop orders in QCD. Equation (\ref{gluequa})
indicates a non-Goldstone mechanism  of chiral
$SU_{\rm L}(N_{\rm F})\times SU_{\rm R}(N_{\rm F})$ symmetry
breaking (for details see~\cite{eur}). This way of violation of chiral
symmetry is unusual but, as it will be discussed further below,
is not in conflict with
the observable meson properties~\cite{bur}.

Manifestations of this quark-gluon dynamics in the spectrum
of collective mesonic excitations were studied in~\cite{efned,bur}
within the model of induced nonlocal quark currents.
Technically, the model is based on
 the bosonization of one-gluon exchange interaction
of quark currents in the presence of the homogeneous (anti-)self-dual
vacuum gluon field (\ref{field}).

The vacuum field
determines basic properties of meson spectrum in the way that is
completely consistent
with experimental data. Namely, due to the zero modes in the
quark propagator the masses of
light pseudoscalar and vector mesons
are strongly split, scalar  and axial mesons
as a simple $q\bar q$ systems in the ground state are absent in the spectrum
(there are no scalar and axial analogies of pions and $\rho$-mesons).
Zero modes affect drastically the pion weak decay
and determine the correct value of constant $f_\pi$.
An entireness of quark and gluon propagators
results in the Regge character of the spectrum of orbital and
radial excitations of light mesons.
Moreover, scalar and axial mesons with appropriate
masses (e.g., $a_1$ meson)
appear in the hyper-fine structure
of orbital excitations of vector mesons.
The mass of heavy quarkonium tends to be equal
to sum of the masses of quarks, the heavy-light meson mass
approaches a mass of a heavy quark, and the weak decay constant for
pseudoscalar heavy-light mesons has asymptotic behaviour
$1/\sqrt{m_Q}$, which is consistent with the Isgur-Wise symmetry.
Quantitatively,
the masses and decay constants of mesons from all different regions of
the spectrum
are described within ten percent inaccuracy.
These different phenomena are displayed with the minimal set
of parameters: gauge coupling constant, strength of the
vacuum field and the quark masses.

This paper proceeds a systematic analysis of possible manifestations
of the vacuum field (\ref{field}) in meson properties.
We intend to give a kind of unified description of
a wide class of
static (masses, decay widths) and dynamical (form factors)
characteristics of different mesons with the set of parameters, that is
minimal for QCD.
Using the values of parameters fitted from the
meson spectrum we got the
electromagnetic form factors of pions to within $15\%$ inaccuracy.
 .

Our main goal
is to study the role of quark zero modes,
induced by an interaction of quark spin with the vacuum field, in the
two-photon decay and transition form factor of $\pi^0$.
The main result is an
observation that the vacuum field (\ref{field}) can be responsible
for both of them. Via zero modes the vacuum field affects in a crucial
manner the form factor and decay constant.
In particular, we show that the spin-field interaction
generates the triangle anomaly in the pion decay to two photons
and leads to
quantitatively satisfactory description
of the transition form factor and decay width.

The main effect of the vacuum field under consideration
in the charge form factor is
an increasing of the contribution of triangle diagram to the form
factor at moderately large transfer momentum $Q^2$.
This effect is due to
the presence of the background field both in the quark propagators and
nonlocal meson-quark vertices, which causes
a specific interplay of
translation and color gauge invariance in the quark loops.
Namely, the translation $x\to x+a$ causes a shift in the
field $B_{\mu\nu}x_\nu \to B_{\mu\nu}x_\nu + B_{\mu\nu}a_\nu$,
which can be compensated by an appropriate gauge transformation.
Therefore, only gauge invariant quantities turn out to be translation
invariant. This results in
nonconservation of energy-momentum in the separate vertices
of Feynman diagrams, but the conservation is still held for a whole diagram
if it is gauge invariant. This peculiarity changes an asymptotic behaviour
of some Feynman diagrams at large momentum transfer.
This is particularly relevant to the triangle diagram for the
charge form factor. Usually the asymptotic
behaviour $\sim (1/Q^2)^{-2}$ of this diagram is a direct consequence
of the behaviour of a meson-quark vertex and quark propagator at short
distances~\cite{roberts}. The presence of the vacuum field changes
the asymptotics cardinally, and it becomes $\sim 1/(Q^2)^{1+m^2_q/B}$,
where $m_q$ is the quark mass and $B$ is the strength of the vacuum field.
This improves the form factor at intermediate and moderately large $Q^2$.
However, the absolute asymptotics of the form factor
within the model under consideration comes from the so-called
bubble diagrams.
These diagrams can be treated as describing the hard rescattering
of quarks inside a pion via virtual gluon exchange.
This mechanism is in agreement with the general analysis within quark
counting rules and perturbative QCD~\cite{radyush,matv}.

In the next section we review shortly the model and introduce the
electromagnetic  interactions into its structure.
Section~III is devoted to calculation of transition form factor, where we
analyze the role of quark zero modes.
Charged form factor of pion is calculated and the
manifestations of the vacuum field in asymptotics of Feynman diagrams
at large momentum transfer is discussed
in sect.~IV.
Several Appendices contain details of calculations.

\section{The Model of Induced Nonlocal Quark Currents}
\subsection{Basic Approximations and Notation}

The functional integral for Euclidean QCD in the presence of the vacuum
field (\ref{field}) can be written in the form~\cite{eur}
\begin{eqnarray}
\label{gf1}
&&Z=
N\int_{\Sigma_B}d\sigma_B
\int_{{\cal G}}D\mu_{G}(G,B)
\int_{\cal F}\prod\limits_{f}^{N_f}Dq_f D\bar q_f
\nonumber\\
&&\exp\left\{\int d^4x
\bar q_f(x)
\left[i\hat\nabla-m_f+g\hat G\right]q_f(x)\right\},
\nonumber\\
&&D\mu_{G}(G,B)=DG
\Delta_{\rm FP}[G,B]
\delta\left[\nabla(B)G\right]
\exp\left\{\int d^4x
{\cal L}_{\rm YM}[G+B]\right\},
\\
&& \hat G=\gamma_\mu G_\mu, \ \
\hat\nabla=\gamma_\mu\nabla_\mu, \ \
\nabla_\mu=\partial_\mu-iB_\mu.
\nonumber
\end{eqnarray}
The vacuum field $B_\mu$ ($gB\equiv B$) is characterized by
angle $\xi$ and two spherical angles
$(\varphi,\theta)$ defining the directions in the color and Euclidean spaces
respectively. The measure of integration $d\sigma_B$ is defined as
\begin{equation}
\label{sigma-vac}
\int_{\Sigma_B} d\sigma_B=\sum_{\pm}
\frac{1}{(4\pi)^2}
\int\limits_{0}^{\pi}{\rm d}\theta \sin\theta
\int\limits_{0}^{2\pi}{\rm d}\varphi
\int\limits_{0}^{2\pi}{\rm d}\xi,
\end{equation}
where the symbol $\sum_{\pm}$ denotes averaging over the self- and
anti-self-dual configurations.

The functional space ${\cal G}$
contains the quantum gauge fields $G_\mu^a$ vanishing
at the space-time infinity.
The fermionic functional integral spans the
Grassmann algebra ${\cal F}$ of square integrable quark fields.
An anti-hermitean representation for Euclidean $\gamma$-matrices is adopted.

The functional integral (\ref{gf1}) contains the intrinsic
dimensionful quantity $B$,
which is a gauge and renormalization invariant strength of the vacuum field,
which by assumption minimizes the QCD effective potential~(e.g., see \cite{mink,leutw,eliz}).
This field strength (or gluon condensate)
provides the natural reference scale for running quark masses
$\bar m_f(\mu)$ and gauge coupling constant $\bar\alpha_s(\mu)$.
Therefore, the strength of the vacuum field $B$,
the quark masses and coupling constant at
the scale $\mu=\sqrt{B}$ can be considered as the physical
(intrinsic) parameters of QCD in representation (\ref{gf1}).
The strength $B$ can be related to the fundamental scale
$\Lambda_{\rm QCD}$~\cite{mink,eur}.
Values of the parameters have to be extracted from the analysis of
hadron spectrum. To be able to do this one needs to
rewrite representation (\ref{gf1}) in terms of composite hadron fields,
describing collective colorless excitations in the QCD vacuum.
For meson fields this program has been realized in
papers~\cite{efned,bur}, where the model of bosonization, referred
below as the model of induced nonlocal quark currents,
has been developed, and the masses and weak decay constants of
mesons from the different regions of meson spectrum have been calculated.
The values of parameters and the results for meson masses are summarized in
Table~\ref{par} and Fig.~1.

A derivation of the model goes through the following steps.
Integrating over gluon fields $G$ one can rewrite Eq.~(\ref{gf1}) in the
form
\begin{eqnarray}
\label{gen2}
&&Z=
N\int_{\Sigma_B}d\sigma_B
\int_{\cal F}\prod\limits_{f}^{N_f}Dq_f D\bar q_f
\nonumber\\
&&\exp\left\{\int d^4x
\bar q_f(x)
\left(i\hat\nabla-m_f\right)q_f(x)
+\sum_{n=2}^{\infty}L_n\right\},
\\
&&L_n=\frac{g^n}{n!}\int d^4y_1...\int d^4y_n
j_{\mu _1}^{a_1}(y_1)\cdot ...
j_{\mu _n}^{a_n}(y_n)
G_{\mu _1...\mu_n}^{a_1...a_n}(y_1,...,y_n\mid B),
 \nonumber\\
&&j_\mu^a(y)=\sum_{f}^{N_F}\bar q_f(x)\gamma_\mu t^aq_f(x).
\end{eqnarray}
The functions $G_{\mu _1...\mu_n}^{a_1...a_n}$ are the exact
$n$-point gluon Green's functions in the external field $B_\mu ^a$.

The model of induced nonlocal quark currents
is based on the assumption that the four-quark interaction $L_2$
in Eq.~(\ref{gen2}) plays the main role in the formation
of mesonic $q\bar q$ collective modes, but the rest of
terms $L_n$ ($n>2$) can be omitted in the first approximation.
Thus, we consider
Eq.~(\ref{gen2}) with the quark-quark interaction
truncated up to the term $L_2$
\begin{eqnarray}
\label{gen3}
Z=N\int_{\Sigma_B}d\sigma_B
\int_{\cal F}\prod\limits_{f}^{N_f}Dq_f D\bar q_f
\exp\left\{\int d^4x
\bar q_f(x)
\left(i\hat\nabla-m_f\right)q_f(x)+L_2\right\}
\nonumber\\
L_2=\frac{g^2}{2}\int\!\!\!\int d^4x d^4y~j_\mu^a(x)
G_{\mu\nu}^{ab}(x,y\mid B)j_\nu^b(y).
\end{eqnarray}
Gluon two-point function $G_{\mu\nu}^{ab}(x,y\mid B)$ is approximated
by the gluon propagator in the external field (\ref{field}), so that
the vacuum field is taken into account exactly but the radiative corrections
are neglected in the model.
The subsequent steps are straightforward and consist
in a standard Fiertz transformation,
decomposition
of $L_2$ into an infinite series
of current-current interaction terms
and  bosonization of the effective
four-quark interactions. For details we refer to papers~\cite{efned,bur}.

The starting point for incorporating the electromagnetic interactions
into the model under consideration is the representation~\cite{bur}
\begin{eqnarray}
\label{l2l}
L_2=\sum_{aJ\ell n}\sum_{j=|\ell-s_J|}^{\ell+s_J}
\frac{1}{2\Lambda^2}G_{J\ell n}^2
\int d^4x\left[{\cal I}^{aJ\ell nj}(x)\right]^2,
\end{eqnarray}
where
\begin{eqnarray}
&& a=0,1,...,N_f,\quad J=S,P,V,A,\quad \ell=0,1,..., \quad n=0,1,...,
\nonumber\\
&& s_{P}=s_{S}=0, \quad s_{V}=s_{A}=1,
\nonumber\\
\label{couplconst}
&&G_{J\ell n}^2=C_J~g^2~
\frac{(\ell+1)}{2^{\ell}n!(\ell +n)!}, \quad C_S=C_P=1/9,
\quad C_V=C_A=1/18
\end{eqnarray}
The currents ${\cal I}^{aJ\ell nj}$ are nonlocal, carry
a complete set of quantum numbers (isospin $a$,
spin-parity in the ground state $J$,
orbital $\ell$
and total angular $j$ momenta, and a radial number $n$)
and has the following form
\begin{eqnarray}
&&{\cal I}_{\mu_1...\mu_\ell}^{bJ\ell n\ell}=
{\cal J}_{\mu_1...\mu_\ell}^{bJ\ell n} \quad {\rm for} \quad J=S,P,
\quad \ell\ge0 \quad (j=\ell),
\nonumber\\
&&{\cal I}_{\alpha}^{bJ0n1}={\cal J}_{\alpha}^{bJ0n}
\quad {\rm for} \quad
J=V,A,  \quad \ell=0  \quad (j=1),
\nonumber
\end{eqnarray}
\begin{equation}
\label{i1}
{\cal I}^{bJ\ell nj}_{\alpha\mu_1...\mu_\ell}= \left\{
\begin{array}{lll}
{1\over(\ell+1)^2}{\cal P}_{\alpha\mu_1...\mu_\ell}\left[
\delta_{\alpha\mu_1}{\cal J}_{\rho,\rho\mu_2...\mu_\ell}^{bJ\ell n}
\right],
~j=\ell-1,\\
{1\over\ell+1}\sum_{i=1}^\ell\left[
{\cal J}_{\alpha,\mu_i...\mu_{i-1}\mu_{i+1}...\mu_\ell}^{bJ\ell n}-
{\cal J}_{\mu_i,\alpha...\mu_{i-1}\mu_{i+1}...\mu_\ell}^{bJ\ell n}\right]
,~j=\ell,\\
{1\over\ell+1}{\cal P}_{\alpha\mu_1...\mu_\ell}\left[
{\cal J}_{\alpha,\mu_1...\mu_\ell}^{bJ\ell n}-
{1\over\ell+1}
\delta_{\alpha\mu_1}{\cal J}_{\rho,\rho\mu_2...\mu_\ell}^{bJ\ell n}
\right],
~j=\ell+1.
\end{array}\right.
\end{equation}
Symbol ${\cal P}_{\alpha\mu_1...\mu_\ell}$  denotes a cyclic
permutation of the indices $(\alpha\mu_1...\mu_\ell)$.
The currents ${\cal J}^{bJ\ell n}$ in Eq.~(\ref{i1}) read
\begin{eqnarray}
\label{cur3}
&&{\cal J}_{\alpha,\mu_1...\mu_\ell}^{bJ\ell n}(x)=
\bar q_f(x)
\left[V_{\mu_1...\mu_\ell}^{bJ\ell n}(x)\right]_{ff^\prime}
q_{f^\prime}(x),
\nonumber\\
\label{vertex2}
&&\left[V_{\alpha,\mu_1...\mu_\ell}^{bJ\ell n}(x)\right]_{ff^\prime}
=M_{ff^\prime}^b\Gamma_\alpha^J\left\{\!\!\left\{F_{n\ell}
\left(
\frac{\stackrel{\leftrightarrow}{\nabla}_{ff^\prime}^2(x)}{\Lambda^2}
\right)
T_{\mu_1...\mu_\ell}^{(\ell)}\left(\frac{1}{i}
\frac{\stackrel{\leftrightarrow}{\nabla}_{ff^\prime}(x)}{\Lambda}\right)
\right\}\!\!\right\},
\nonumber\\
\label{ffactor}
&& F_{n\ell}(4s)=s^n~\int_0^1 dtt^{\ell+n}e^{st},
\\
&&\stackrel{\leftrightarrow}{\nabla}_{ff^\prime}=
\xi_f\stackrel{\leftarrow}{\nabla} -
\xi_{f^\prime}\stackrel{\rightarrow}{\nabla}, \ \
\stackrel{\rightarrow}{\nabla}=\stackrel{\rightarrow}{\partial}-iB, \ \
\stackrel{\leftarrow}{\nabla}=\stackrel{\leftarrow}{\partial}+iB, \ \
 \xi_f=m_f/(m_f+m_{f^\prime}),
\nonumber\\
&&\Gamma_S=I, \quad \Gamma_P=i\gamma_5, \quad \Gamma_V^\alpha=\gamma^\alpha,
\quad \Gamma_A^\alpha=\gamma_5\gamma^\alpha.
\nonumber
\end{eqnarray}
The polynomials $T_{\mu_1...\mu_\ell}^{(\ell)}(x)$ are the irreducible tensors
of $O(4)$.
The doubled brackets in
Eq.~(\ref{vertex2}) mean that the covariant derivatives commute
inside these brackets. For the sake of simplicity the scale
$\Lambda^2=\sqrt{3}B$ is introduced and a particular direction
$n^a=\delta^{a8}$ of
the vacuum field in the color space is fixed, so that:
\begin{eqnarray}
\label{vv}
\hat B_{\mu\rho}\hat B_{\rho\nu}=-v^2\Lambda^4\delta_{\mu\nu},
\ v={\rm diag}(1/3,1/3,2/3).
\end{eqnarray}
Further details concerning above representations can be found in
papers~\cite{efned,bur}.

Form-factors $F_{n\ell}(s)$ appeared as a result
of decomposition of bilocal quark currents over the  orthogonal
complete set of polynomials for which gluon propagator plays the role of a
weight function.
The are entire
analytical functions in the complex $s$-plane, which is a manifestation
of the gluon confinement.
Now an interaction
between quarks is expressed in terms of the nonlocal quark currents,
which are elementary currents of the system in the sense of
classification over quantum numbers.

\subsection{Electromagnetic Interactions}

Interaction of quarks with electromagnetic field $A_\mu$
can be introduced into
representation (\ref{gen3}) with $L_2$ written in the form of
Eq.~(\ref{l2l})
by means of the minimal substitution (see also~\cite{gross})
\begin{eqnarray}
\label{min}
&&\stackrel{\rightarrow}{\nabla}\to
\stackrel{\rightarrow}{\nabla}-ie_f A(x),
\nonumber \\
&&\stackrel{\leftarrow}{\nabla}\to
\stackrel{\leftarrow}{\nabla}+ie_f A(x)
\end{eqnarray}
both in the free quark Lagrangian and vertices $V_{aJ\ell nj}$.
Here $e_f=e{\cal Q}_f$, and $\cal Q$ is the charge matrix.

Equation (\ref{gen3}) takes the form (${\cal N}=\{aJ\ell nj\}$)
\begin{eqnarray}
\label{gen4}
&&Z[A]=\int d\sigma_{\rm B}\int DqD\bar q
\exp\biggl\{\int{\rm d}^4x \bar q_f(x)
[i\gamma_\mu\stackrel{\rightarrow}{\nabla}_\mu-m_f+e_f\gamma_\mu
A_\mu(x)]q_f(x)
\nonumber \\
&&+\sum_{\cal N}\frac{1}{2\Lambda^2}G^2_{\cal N}\int{\rm d}^4x
\left|{\cal I}_{\cal N}(x\mid A)-{\rm Tr}
V_{\cal N}(x\mid 0)S(x,x\mid 0)\right|^2\biggr\},
\end{eqnarray}
where the currents ${\cal I}_{\cal N}(x\mid A)$ include electromagnetic
field $A$ through the vertex functions
\begin{eqnarray}
\label{vertex3}
&&V^{bJ\ell n}(x\mid A)=
M^b\Gamma^J\left\{\!\!\left\{F_{n\ell}
\left(\frac{\stackrel{\leftrightarrow}{D}^2(x)}{\Lambda^2}\right)
T^{(\ell)}\left(\frac{1}{i}
\frac{\stackrel{\leftrightarrow}{D}(x)}{\Lambda}\right)
\right\}\!\!\right\}, \\
\label{D}
&&\stackrel{\leftrightarrow}{D}^\mu_{ff^\prime}=
\xi_f\left[\stackrel{\leftarrow}{\nabla}^\mu+ie_fA^\mu(x)\right]-
\xi_{f^\prime}\left[\stackrel{\rightarrow}{\nabla}^\mu
-ie_{f^\prime}A^\mu(x)\right],
\end{eqnarray}
One can check that under the gauge transformation
\begin{eqnarray}
\label{gauge}
&&q_f(x)\to q_f^\prime(x)={\rm e}^{ie_f\alpha(x)}q_f(x),
\nonumber \\
&&\bar q_f(x)\to \bar q_f^\prime(x)=\bar q_f(x){\rm e}^{-ie_f\alpha(x)},
\nonumber\\
&&A_\mu(x)\to A^{\prime}_\mu(x)=A_\mu(x)+\partial_\mu\alpha(x)
\end{eqnarray}
the currents $\cal I_N$ are transformed
with the electric
charge $(e_f-e_{f^\prime})$ corresponding to a meson composed of
$q_f$ and $\bar q_{f^\prime}$:
\begin{eqnarray}
\label{gcur}
&&{\cal I}_{\cal N}(x\mid A)\to {\cal I}_{\cal N}^{\prime}(x\mid A)
={\rm e}^{i(e_f-e_{f^\prime})\alpha(x)}{\cal I}_{\cal N}(x\mid A),
\nonumber \\
&&{\cal I}_{\cal N}^\dagger(x\mid A)\to
{\cal I}_{\cal N}^{\prime_\dagger}(x\mid A)
={\cal I}_{\cal N}^\dagger(x\mid A){\rm e}^{-i(e_f-e_{f^\prime})\alpha(x)}.
\end{eqnarray}
The  current-current interaction
${\cal I}^\dagger_{\cal N}(x){\cal I}_{\cal N}(x)$
is invariant under $U(1)$
transformations (\ref{gauge}).

By means of the standard bosonization procedure~\cite{efned,bur}
applied to Eq.~(\ref{gen4}) the functional integral $Z[A]$ can be
represented in terms of the composite meson fields $\Phi_{\cal N}$
\begin{eqnarray}
\label{genlast}
&&Z[A]=N\int \prod_{\cal N}D\Phi_{\cal N}
\exp\left\{\frac{1}{2}\int\!\!\int {\rm d}^4x{\rm d}^4y
\Phi_{\cal N}(x)\left[\left(
\Box-M_{\cal N}^2\right)\delta(x-y)\right.\right. \nonumber\\
&&-\left.\left.h_{\cal N}^2\Pi_{\cal N}^R(x-y)\right]
\Phi_{\cal N}(y)+I_{\rm int}\left[\Phi\mid A\right]\right\},
\end{eqnarray}
\begin{eqnarray}
&&I_{\rm int}=
-\int d^4x h_{\cal N}
\Phi_{\cal N}(x)\left[
\Gamma_{\cal N}(x\mid A)-
\Gamma_{\cal N}(x\mid 0)
\right]
\nonumber\\
&&-\frac{1}{2}\int d^4x_1\int d^4x_2 h_{\cal N}h_{\cal N^\prime}
\Phi_{\cal N}(x_1)\left[\Gamma_{\cal N N'}(x_1,x_2\mid A)-
\delta_{\cal NN'}\Pi_{\cal N}(x_1-x_2)\right]\Phi_{\cal N'}(x_2)
\nonumber\\
&&-\sum_{m=3}\frac{1}{m}\int d^4x_1...\int d^4x_m
\prod_{k=1}^mh_{{\cal N}_k}\Phi_{{\cal N}_k}(x_k)
\Gamma_{{\cal N}_1...{\cal N}_m}(x_1,...,x_m\mid A),\nonumber\\
\label{Gamma}
&&\Gamma_{{\cal N}_1...{\cal N}_m}=\int d\sigma_{\rm B}
{\rm Tr}\left\{V_{{\cal N}_1}(x_1\mid A)S(x_1,x_2\mid A)...
V_{{\cal N}_m}(x_m\mid A)S(x_m,x_1\mid A)\right\}.
\end{eqnarray}
At zero electromagnetic field the effective meson action in
Eq.~(\ref{genlast})
coincides with the action derived in~\cite{bur}.
Meson masses $M_{\cal N}$ are calculated by solving the equations
\begin{eqnarray}
\Lambda^2+G^2_{\cal N}\tilde\Pi_{\cal N}(-M^2_{\cal N})=0,
\nonumber
\end{eqnarray}
where $\tilde\Pi_{\cal N}(p^2)$ is a diagonal part of the Fourier
transform of the two-point function $\Gamma_{{\cal N}{\cal N}^\prime}$
at zero electromagnetic field.

The meson-quark coupling constants are defined by the relations:
\begin{eqnarray}
\label{h}
h^2_{\cal N}=
1/\tilde\Pi^\prime_{\cal N}(p^2)|_{p^2=-M^2_{\cal N}}.
\end{eqnarray}

The integral over directions of the field and the sum
over self- and anti-self-dual configurations
has been transferred to the exponent in Eq.~(\ref{genlast})
and included into the vertices $\Gamma$, Eq.~(\ref{Gamma}).
This assumes that the vacuum field configurations
in the different (separated by meson lines) quark loops
are considered as
independent from each other.
Such a prescription should
reflect effectively  the domain structures in the vacuum.

To get various meson-photon amplitudes one has to decompose the vertex
functionals $\Gamma_{{\cal N}_1...{\cal N}_m}$
into a series over the electromagnetic field $A$, as is shown schematically
in Fig.~2.
This is achieved by a decomposition of the quark propagators,
\begin{eqnarray}
\label{s-decomp}
S_f(x,y|A)&=&S_f(x,y)
\nonumber\\
&+&\sum\limits_{n=1}^\infty e^n\int d^4z_1...\int d^4z_n
S_f(x,z_1)\gamma_{\mu_1}A_{\mu_1}(z_1)...
\gamma_{\mu_n}A_{\mu_n}(z_n)S_f(z_n,y),
\end{eqnarray}
and vertices
\begin{eqnarray}
\label{v-decomp}
V_{\cal N}(x|A)&=&V_{\cal N}(x)
\nonumber\\
&+&\sum\limits_{n=1}^\infty e^n\int d^4z_1...
\int d^4z_n V^{(n)}_{{\cal N};\mu_1...\mu_n}(x,z_1,...,z_n)
A_{\mu_1}(z_1)...A_{\mu_n}(z_n).
\end{eqnarray}
A regular procedure for calculation of the vertices
$V^{(n)}_{{\cal N};\mu_1...\mu_n}$ is described in Appendix~A.

The lowest vertex for charged pion, the case of particular
importance for further calculations, comes from
the current ${\cal J}_{\pi^{+}}$, Eqs.~(\ref{cur3}) and (\ref{vertex3}),
\begin{eqnarray}
\label{curr}
&&{\cal J}_{\pi^{+}}(x)=
\bar d(x)i\gamma_5 F_{00}
\left(\frac{\stackrel{\leftrightarrow}{D}^2(x)}{\Lambda^2}\right)u(x)
=\bar d(x)i\gamma_5 F_{00}\left(\frac{\stackrel
{\leftrightarrow}{\nabla}^2(x)}{\Lambda^2}\right)u(x)
\nonumber \\
&&+i\gamma_5e\int d^4y \bar d(x)A_\mu(y)\Gamma_\mu(x,y)u(x)+O(e^2),
\end{eqnarray}
where
\begin{eqnarray}
\label{finalb}
\Gamma_\mu(x,y)&=&
\int\frac{d^4q}{(2\pi)^4}{\rm e}^{iq(x-y)}
\int\limits_0^1\int\limits_0^1dt d\beta
\frac{t}{4\Lambda^2}
\left[2i\stackrel{\leftrightarrow}{\nabla}_\mu(x)-q_\mu\right]
\nonumber \\
&&\times\exp\left\{\frac{t}{4\Lambda^2}
\left[\stackrel{\leftrightarrow}{\nabla}^2(x)
+\beta (2i\stackrel{\leftrightarrow}{\nabla}(x)-q)q\right]\right\}.
\end{eqnarray}
The first term in Eq.~(\ref{curr}) is the nonlocal quark
current
in the absence of field $A_\mu$
but the second term describes an
interaction of photon with a quark inside pion which is
characterized by the form factor $\Gamma_\mu(x,y)$~\cite{gross}.

The quark propagator in the homogeneous (anti-)self-dual gluon field
$S_f(x,y)$ is a solution to equation
\begin{eqnarray}
&&(i\gamma_\mu\nabla_\mu-m_f)S_f(x,y)=-\delta(x-y),
\nonumber\\
\label{prop}
&&S_f(x,y)=e^{\frac{i}{2}x_\mu\hat B_{\mu\nu}y_\nu}H_f(x-y),
\nonumber\\
&&H_f(z)=\left(i\gamma_\mu\nabla_\mu+m_f\right)
\left(-\nabla^2+m_f^2-\sigma_{\alpha\beta}
\hat B_{\alpha\beta}\right)^{-1}\delta(z).
\end{eqnarray}
The term $\sigma_{\alpha\beta}\hat B_{\alpha\beta}$
in Eq.~(\ref{prop}) is particularly important. It describes an
interaction of a quark spin with the vacuum field and is responsible
for the quark zero modes.
Fourier transform of the translation invariant part $H_f$
reads~\cite{bur,eur}
\begin{eqnarray}
\label{qsol}
\tilde H_f(p)=
\frac{1}{2v\Lambda^2}\int\limits_{0}^{1}ds
e^{-\frac{p^2}{2v\Lambda^2}s}\left(\frac{1-s}{1+s}\right)
^{\frac{m_f^2}{4v\Lambda^2}}
\left[
p_\alpha\gamma_\alpha\pm is\gamma_5\gamma_\alpha f_{\alpha\beta}p_\beta
\right.
\nonumber \\
\left.
+m_f\left(P_\pm +P_\mp
\frac{1+s^2}{1-s^2}-\frac{i}{2}\gamma_\alpha f_{\alpha\beta}\gamma_\beta
\frac{s}{1-s^2}\right)
\right],\\
f_{\mu\nu}=\frac{\lambda^8}{v\Lambda^2}B_{\mu\nu},\nonumber
\end{eqnarray}
where $P_\pm=(I\pm\gamma_5)/2$ and the upper (lower) sign corresponds to
(anti-)self-dual configuration.

Propagator (\ref{qsol}) is an entire
analytical function in the complex momentum plane, which is treated
as quark confinement.
The term $\sigma_{\alpha\beta}\hat B_{\alpha\beta}$
in Eq.~(\ref{prop}) is particularly important. It describes an
interaction of a quark spin with the vacuum field and is responsible
for the quark zero modes. Contribution of zero modes to the propagator
is seen in Eq.~(\ref{qsol}) as a singilarity of the integrand
at $s=1$ which is integrable unless $m_f=0$.

Below we show that the spin-field interaction is of crucial importance
for the transition form factor of neutral pion. The presence
of the vacuum field in the phase factor in quark propagator and
in covariant derivatives in the vertices  turns out to be important for
the charge form factor.

\section{ Decay $\pi^0\to \gamma\gamma$
and  $\gamma^{\ast}\pi^0\to\gamma$ Transition Form Factor}

A vertex relevant to the interaction of neutral pion
 and two photons with momenta $p$, $k_1$ and $k_2$, correspondingly,
has the following general structure
\begin{eqnarray}
\label{1}
&&T_\pi^{\mu\nu}(p,k_1,k_2)=i\delta^{(4)}(k_1+k_2-p)
\epsilon^{\mu\nu\alpha\beta}
k_1^{\alpha}k_2^{\beta}T_{\pi}(p^2,k_1^2,k_2^2),
\end{eqnarray}
where $T_{\pi}(p^2,k_1^2,k_2^2)$ is a scalar function.
It should be remembered that we have started with the Euclidean formulation
of QCD, and $k_1$, $k_2$ and $p$ are Euclidean momenta.

For $\gamma^{\ast}\pi^0\to\gamma$ transition the final photon
$\gamma$ is on the mass shell $k_2^2=0$, whereas $k_1^2=Q^2>0$
for the virtual photon $\gamma^\ast$.
The transition form factor
$F_{\gamma\pi}(Q^2)$
is then defined as
\begin{eqnarray}
\label{2}
&&F_{\gamma\pi}(Q^2)=T_\pi(-M^2_\pi,Q^2,0).
\end{eqnarray}
The two-photon decay coupling constant
\begin{eqnarray}
\label{3}
g_{\pi\gamma\gamma}=T_\pi(-M^2_\pi,0,0)
\end{eqnarray}
and the decay width
\begin{eqnarray}
\label{4}
&&\Gamma(\pi^0\to\gamma\gamma)=\frac{\pi}{4}\alpha^{2}M^3_\pi
g_{\pi\gamma\gamma}^2.
\end{eqnarray}
correspond to kinematics with on-shell photons and pion:
$k_1^2=k_2^2=0$ and $p^2=-M_\pi^2$.

As follows from the effective action in Eq.~(\ref{Gamma}),
the one loop approximation of  the
vertex (\ref{1}) is described by the
triangle diagram shown in Fig.~3.  Other
one loop diagrams with
photon lines attached to the meson-quark vertex are identically equal to zero
due to the parity conservation.

As has been mentioned above,
the interaction of quark spin with the homogeneous vacuum field
generates an infinite number of zero modes of Dirac operator and
leads to a nonzero quark condensate density, which indicates
a breakdown of chiral symmetry by the vacuum gluon field \cite{eur}.
A contribution of zero modes to the polarization diagrams
of light mesons is responsible for mass splitting between vector and
pseudoscalar mesons and smallness of pion mass~\cite{bur}.
The main goal of this section is to show that the zero modes
generate the triangle anomaly and
play the decisive role in the two-photon decay of $\pi^0$
and transition form factor $F_{\gamma\pi}$.
To demonstrate this qualitatively it suffices to consider
the triangle diagram (Fig.~3) in the limit of vanishing quark masses
$m_d=m_u\to0$.

The quark propagator in the external (anti-)self-dual field~(\ref{field})
has the following standard representation
in terms of the matrix elements of the projection
operators ${\cal P}_n$ onto the subspaces corresponding to
different eigen numbers $\lambda_n$ of
Dirac operator in the external field~\cite{brown1,eur}
$$
S(x,y)=\sum\limits_{n=0}^{\infty}
\frac{{\cal P}_n(x,y)}{m+i\lambda_n}.
$$
One can separate the contribution of the zero eigenmodes
and normal modes to the propagator
\begin{eqnarray}
&&S(x,y)=S^\prime(x,y)+S_0(x,y),
\nonumber\\
\label{qpropn1}
&&S^\prime(x,y)=\stackrel{\rightarrow}{i\hat\nabla_x}
\Delta(x,y)P_{\pm}+
\Delta(x,y)\stackrel{\leftarrow}{i\hat\nabla_y}P_{\mp}
+O(m),\\
&&\stackrel{\rightarrow}{\nabla}=\stackrel{\rightarrow}{\partial}
-i B, \ \
\stackrel{\leftarrow}{\nabla}=\stackrel{\leftarrow}{\partial}
+i B,
\nonumber\\
&& S_0(x,y)={\cal P}_0(x,y)/m.
\nonumber
\end{eqnarray}
Here
$\Delta(x,y)$ is the scalar massless propagator
in the background field (\ref{field}),
and $P_\pm=(1\pm\gamma_5)/2$.
The projector onto the zero mode subspace looks as
\begin{eqnarray}
\label{p0}
&&{\cal P}_0=\frac{n^2B^2}{\pi^2}f(x,y)P_{\mp}\Sigma_{\mp},
\\
&&
\Sigma_{\pm}=\frac{1}{2}\left(1\pm\Sigma_j b_j\right), \
\Sigma_i=\frac{1}{2}\varepsilon_{ijk}\sigma_{jk}, \
\sigma_{jk}=[\gamma_j,\gamma_k]/2i,
\nonumber\\
&&b_j=B_j/B, \ B_i=-\frac{1}{2}\varepsilon_{ijk}B_{jk}, \
i,j,k=1,2,3
\nonumber\\
&&f(x,y)=\exp\left\{
-\frac{1}{2}\sqrt{n^2}B(x-y)^2+
inx_\mu B_{\mu\nu}y_\nu\right\},\\
&&\int d^4z{\cal P}_0(x,z){\cal P}_0(z,y)={\cal P}_0(x,y).
\nonumber
\end{eqnarray}
The projector
${\cal P}_0(x,y)$
contains the projection matrices $P_{-}$
(self-dual field) or $P_+$ (anti-self-dual field) and
$\Sigma_-$ ($n>0$) or $\Sigma_+$ ($n<0$).
The matrix $\Sigma_+$ ($\Sigma_-$) can be seen as the projector onto
the quark state with the spin orientated along (against)
the chromomagnetic field $B_j$.
Now consider the $n$-point quark loop
\begin{equation}
\label{loops}
{\rm Tr}\Gamma_1(S_0(x_1,x_2)+S^\prime(x_1,x_2))
\Gamma_2 (S_0(x_2,x_3)+S^\prime(x_2,x_3))
\dots
\Gamma_n (S_0(x_n,x_1)+S^\prime(x_n,x_1)),
\end{equation}
where $\Gamma_k$ are some Dirac matrices. Chiral structure
of zero mode part ${\cal P}_0$, Eq.~(\ref{p0}),
and normal mode part $S^\prime$, Eq.~(\ref{qpropn1}),
suggests that
the loops with all vector vertices are regular in the massless limit
(for details see~\cite{eur}), while diagrams with one pseudoscalar
and $n-1$ vector vertices are singular and behaves as $1/m_u$.
The second kind is just the case of Fig.~3.

Thus the triangle diagram for the
amplitude (\ref{1}) is proportional to $1/m_u$  as $m_u\to0$.
However, the amplitude contains also the pion-quark
coupling constant $h_\pi$ defined by the pion polarization
function according to Eq.~(\ref{h}).
The polarization function is simply a two-point quark loop
of the form (\ref{loops}) with  pseudoscalar vertices.
In the massless limit it diverges as $1/m_u^2$ and, hence,
coupling constant $h_\pi\propto m_u$.
Thus the massless limit of the form factor, that is a product of
coupling constant $h_\pi$ and
the quark loop, is nonzero: $\lim_{m_u\to0}F_{\gamma\pi}\not=0$.
This anomaly is determined completely by a
contribution of quark zero modes
to triangle diagram and coupling constant.

Let us consider this important peculiarity in more details.
Using the proper time representation of the quark propagator (\ref{qsol})
and meson-quark vertex (\ref{ffactor}), evaluating the trace of Dirac
matrices,
calculating the loop momentum integrals and averaging over different
configurations of the vacuum field,
we represent the  form factor $F_{\gamma\pi}(Q^2)$
as an  integral over proper times
$t$ and $s_1$, $s_2$, $s_3$, corresponding to the vertex and
propagators, respectively (see Fig.~3):
\begin{eqnarray}
\label{5}
&&F_{\gamma\pi}(Q^2)=\frac{1}{\Lambda}\frac{h_\pi}{2\sqrt{2}\pi^2}
{\rm Tr}_v\int \limits_{0}^{1}...\int \limits_{0}^{1}dtds_1ds_2ds_3
\left[\left(\frac{1-s_1}{1+s_1}\right)
\left(\frac{1-s_2}{1+s_2}\right)
\left(\frac{1-s_3}{1+s_3}\right)\right]^{m^2_u/4v}
\nonumber\\
&&\times
\sum\limits_{i=1,2,3}\frac{m_u}{1-s_i^2}\Phi_i(M^2_\pi,Q^2;s,t)
\exp\left[M^2_\pi\phi(s,t)-Q^2\varphi(s,t)\right],\\
&&\phi(s,t)=(2s_1s_3+vt(s_1+s_3-s_2(1+s_1s_3)))/4v\chi,
\nonumber\\
&&\varphi(s,t)=s_2(s_3+vt(1+s_1s_3))/2v\chi,
\nonumber\\
&&\chi=2vt(1+s_1s_2+s_1s_3+s_2s_3)+(s_1+s_2+s_3+s_1s_2s_3).
\nonumber
\end{eqnarray}
Here and below we use the shorthand notation for
dimensionless  ratios: $Q^2=Q^2/\Lambda^2$,
$m_u=m_u/\Lambda$, $M_\pi=M_\pi/\Lambda$.
The symbol ${\rm Tr}_v$ means summation
over the elements of the diagonal matrix
$v$ written in Eq.~(\ref{vv}).
The details of calculation of $F_{\gamma\pi}(Q^2)$ and an explicit form of
functions $\Phi_i(M^2_\pi,Q^2;s,t)$ can be found in Appendix~B.

The singularities $(1-s_i)^{-1}$ of the integrand in Eq.~(\ref{5})
at $s_i\to 1$ appear from the zero mode contribution
to the quark propagator (see the second line in Eq.~(\ref{qsol})).
These singularities lead to the $1/m_u$-dependence of the integral
in Eq.~(\ref{5}) in the limit $m_u\ll 1$:
\begin{eqnarray}
\label{6}
F_{\gamma\pi}(Q^2)=\frac{h_\pi}{m_u}I(Q^2).
\end{eqnarray}
Here $I$ does not depend on $m_u$.
In the massless limit pion polarization function
$\tilde\Pi_\pi(-M_\pi^2)$ looks as~\cite{bur}
\begin{equation}
\label{pion}
\tilde\Pi_\pi(-M_\pi^2)=
\frac{1}{m_u^2}\frac{16\Lambda^2}{\pi^2M_\pi^4}{\rm Tr}_v v^2
\left[\exp\left\{\frac{M_\pi^2}{8v}\right\}-
\exp\left\{\frac{M_\pi^2}{8v(1+2v)}\right\}\right]^2,
\end{equation}
hence the effective coupling
constant $h_\pi$ in Eq.~(\ref{6}) behaves as
\begin{equation}
\label{hmqq}
h_\pi=1/\sqrt{\tilde\Pi'_\pi(-M_\pi^2)}\sim m_u,
\end{equation}
and $F_{\gamma\pi}(Q^2)$ does not vanish as $m_u\to 0$ but approaches
a constant
value.
The same is valid for decay width (\ref{4}).

Numerical results for the form factor and decay width
are represented in Fig.~4 and Table~II.
The model parameters are given in
Table~I. It should be remembered that we simply use
the values of parameters fixed from the description
of the meson spectrum~\cite{bur}.
The solid curve in Fig.~4 corresponds to Eq.~(\ref{5}).
The radius for $\gamma^{\ast}\pi^0\to\gamma$ transition defined as
\begin{eqnarray}
<{\rm r}^2_{\gamma\pi}>=-6 \frac{F'_{\gamma\pi}(0)}{F_{\gamma\pi}(0)},
\nonumber
\end{eqnarray}
is equal to .57~fm, that have to be compared with
$r_{\gamma\pi}^{\rm exp}=.65\pm.03$~\cite{bebek}.

Just to illustrate the crucial role of the quark zero modes we
show  the form factor calculated with zero mode
contribution eliminated from the quark propagator
(the long dashed line).
Zero modes drastically affect also the two-photon decay constant
$g_{\pi\gamma\gamma}$ and decay width $\Gamma(\pi^0\to\gamma\gamma)$,
as is seen from Table~II.

Above consideration supports the picture of chiral symmetry breaking
due to the fermion zero modes induced by the homogeneous
(anti-)self-dual vacuum gluon field, which has been developed
in our previous papers~\cite{bur} and \cite{eur}.
We can conclude that within the model of induced nonlocal currents
the experimental
data for $\gamma^{\ast}\pi^0\to\gamma$ transition form factor and
two photon decay constant can be
explained by the same reason as
a smallness of pion mass, splitting of the masses
of vector and pseudoscalar light mesons and
weak decay constants of pions and kaons.
A general physical reason is an interaction of a quark spin with
the vacuum homogeneous gluon field. This spin-field interaction
appears to be a dominating effect in the above discussed phenomena.

It is appropriate to mention here, that, possibly,
there is another, completely independent
of our considerations,
manifestation of an interaction of the quark spin with a
long-range gluon field in the QCD vacuum. As is reported in
paper~\cite{efremov} the experimentally observed sign of the jet handedness
correlation can be naturally understood if the jet fragmentation occurs
in the background of vacuum gluon field that is (almost) homogeneous
within some characteristic region. The spin-field interaction
plays here the leading, qualitatively important, role.

The asymptotic behaviour of the Feynman diagrams
in the limit of large momentum transfer is an additional
point where the homogeneous vacuum field could be seen as a relevant
effect. We will pay more attention
to this in the next section, where the charge form factor of pion is
considered.
However, just for comparison, it is advantageous to calculate the asymptotic
form of the transition form factor.

A behaviour of the triangle diagram for
the transition form factor in the limit
$Q^2\gg\Lambda^2$ can be easily estimated.
Equation (\ref{5}) can be rewritten
in the form
\begin{eqnarray}
\label{8}
F_{\gamma\pi}(Q^2/\Lambda^2)=\frac{1}{\Lambda}
\int\limits_{0}^{1}...\int \limits_{0}^{1}dtds_1ds_2ds_3
\Phi(Q^2s_2/\Lambda^2,M_\pi/\Lambda,m_u/\Lambda;t,s_1,s_3)
\nonumber\\
\times\exp\left[-\frac{Q^2}{\Lambda^2}\varphi(s,t)\right],
\nonumber\\
\varphi(s_1,s_2,s_3,t)=s_2
(s_3+vt(1+s_1s_3))/2v\chi(s_1,s_2,s_3,t),
\nonumber
\end{eqnarray}
which just underlines that
the integrand $\Phi$ depends on $Q^2$ and $s_2$ in the combination $Q^2s_2$,
where the variable $s_2$ corresponds to the quark propagator situated
in Fig.~3
between two electromagnetic vertices.
Here we have restored the
dimensional notation for the masses and momentum $Q$.
Function $\chi$ is given in Eq.~(\ref{5}).
For any fixed values of $t,s_1,s_3$, the function $\varphi$  is
increasing in $s_2$ and
gets the lowest value
at the point $s_2=0$.
This corresponds to the ultraviolet regime in the space of integration
variables, that is usual for large momentum
asymptotics~\cite{radyush}.
An integral over $s_2$ can be evaluated by the
Laplace method and, for $Q^2\gg\Lambda^2$,
we arrive at relation
\begin{eqnarray}
\label{10}
&&Q^2F_{\gamma\pi}(Q^2/\Lambda^2)=
C\Lambda + O(\Lambda^2/Q^2)
\approx 0.2 {\rm Gev} + O(\Lambda^2/Q^2) ,
\nonumber\\
&&
C={\rm Tr}_v\int\limits_{0}^{1}...\int\limits_{0}^{1}
dtds_1ds_3\int\limits_{0}^{\infty}ds_2
\Phi(s_2,M_\pi/\Lambda,m_u/\Lambda;t,s_1,s_3)
\nonumber\\
&&\times\exp\left[-s_2(s_3+vt(1+s_1s_3))/2v\chi(s_1,0,s_3,t)
\right].
\nonumber
\end{eqnarray}
This result has to be compared with the Brodsky-Lepage limit~\cite{br}
$$Q^2F_{\gamma\pi}(Q^2) \to 2F_\pi = .186 {\rm Gev}.$$
We see that the asymptotic regime in the form factor is realized in the
usual way consistent with the analysis within factorization hypothesis
and QCD sum rule
approaches~\cite{rad1}.

\section{The pion charge form factor}

According to the effective action~(\ref{Gamma}),
the one-loop amplitude for the
processes $\pi^\pm\gamma^\ast\to\pi^\pm$
is described by the triangle and bubble
diagrams shown in Figs.~5a and 5b correspondingly.
It has the following  structure
\begin{eqnarray}
\label{c1}
&&\Lambda^\mu(k_1,k_2,q)=\delta^{(4)}(k_1-k_2+q)
\left[(k_1+k_2)^\mu F_1(k_1^2,k_2^2,q^2)
+q_\mu F_2(k_1^2,k_2^2,q^2)\right],
\end{eqnarray}
where $k_1$ and $k_2$ are the pion momenta and $q$ is a momentum
of virtual photon $\gamma^\ast$.
As is known,
within the minimal substitution scheme of introduction
of electromagnetic interactions the vertex $\Lambda^\mu$ given by
a sum of triangle and
bubble diagrams satisfies
the Ward-Takahashi identity~\cite{gross,valera}.

The pion charge form factor $F_\pi(Q^2)$ is defined by
the relation
\begin{eqnarray}
\label{c2}
&&F_\pi(Q^2)=F_\pi^\triangle(Q^2)+F_\pi^\circ(Q^2),
\nonumber\\
&&F_\pi^\triangle=F_1^\triangle(-M_\pi^2,-M_\pi^2,Q^2), \
F_\pi^\circ=F_1^\circ(-M_\pi^2,-M_\pi^2,Q^2).
\nonumber
\end{eqnarray}
The details of calculation of contributions of triangle
$F_\pi^\triangle(Q^2)$ and bubble
$F_\pi^\circ(Q^2)$ diagrams to the form factor are relegated to
Appendix~C. The result expressed in the form of the proper time
integrals reads:
\begin{eqnarray}
\label{c3}
F_\pi^\triangle(Q^2)&=&\frac{h_\pi^2}{4\pi^2}
{\rm Tr}_v\int\limits_{0}^{1}...\int \limits_{0}^{1}
dt_1dt_2ds_1ds_2ds_3
\left[\left(\frac{1-s_1}{1+s_1}\right)
\left(\frac{1-s_2}{1+s_2}\right)
\left(\frac{1-s_3}{1+s_3}\right)\right]^{m^2_u/4v}
\nonumber\\
&&\times\left[Q^2\Phi_1(M^2_\pi,Q^2;s,t)+
m_u^2\Phi_2(M^2_\pi,Q^2;s,t)+M_\pi^2\Phi_3(M^2_\pi,Q^2;s,t)
\right.
\nonumber\\
&&\left.
+\Phi_4(M^2_\pi,Q^2;s,t)\right]
\exp\left[M^2_\pi\phi(s,t)-Q^2\varphi(s,t)\right],
\end{eqnarray}
\begin{eqnarray}
&&\phi=(2s_1s_2+2s_1s_3+v(t_1+t_2)
(s_1+s_2+s_3+s_1s_2s_3))/4v\chi,
\nonumber\\
&&\chi=2v(t_1+t_2)(1+s_1s_2+s_1s_3+s_2s_3)
+(1+4v^2t_1t_2)(s_1+s_2+s_3+s_1s_2s_3),
\nonumber\\
&&\varphi=[s_2s_3+vt_1s_2(1+s_1s_3)+vt_2s_3(1+s_1s_2)
+v^2t_1t_2(1+s_1s_2+s_1s_3+s_2s_3)]/2v\chi,
\nonumber
\end{eqnarray}
for triangle diagram, and
\begin{eqnarray}
\label{b}
&&F_\pi^\circ(Q^2)=\frac{h_\pi^2}{4\pi^2}
{\rm Tr}_v\int \limits_{0}^{1}...\int \limits_{0}^{1}
dt_1dt_2ds_1ds_2d\beta
\left[\left(\frac{1-s_1}{1+s_1}\right)
\left(\frac{1-s_2}{1+s_2}\right)
\left(\frac{1-s_3}{1+s_3}\right)\right]^{m^2_u/4v}
\nonumber\\
&&\times\left[\frac{m_u^2}{(1-s_1^2)(1-s_2^2)}
\Phi_1^\circ(M^2_\pi,Q^2;\beta,s,t)+
\Phi_2^\circ(M^2_\pi,Q^2;\beta,s,t)\right]
\nonumber\\
&&\times\exp\left[M^2_\pi\phi^\circ(s,t)
-Q^2\varphi^\circ(\beta,s,t)\right]
\nonumber\\
&&\phi^\circ=[2s_1s_2+v(t_1+t_2)(s_1+s_2)]/4v\chi^\circ,
\nonumber \\
&&\varphi^\circ=
\frac{vt_2\beta}{2v\chi^{\circ}}\left[s_1+vt_1(1+s_1s_2)
+vt_2(1-\beta)[1+s_1s_2+2vt_1(s_1+s_2)]\right],
\nonumber\\
&&\chi^\circ=2v(t_1+t_2)(1+s_1s_2)+(1+4v^2t_1t_2)(s_1+s_2),
\end{eqnarray}
for bubble diagram.
We have used the shorthand dimensionless notation
for momentum $Q$ and masses. Functions $\Phi_i$ and $\Phi^\circ_i$
are written in Appendix~C.

For the parameter values given in Table~1
the charge form factor defined by Eqs.~(\ref{c3}) and (\ref{b})
is plotted in
Fig.~6 by the solid line.
The electromagnetic radius takes the value:
\begin{eqnarray}
\langle r^2_{\pi}\rangle=-6 \frac{F'_{\pi}(0)}{F_{\pi}(0)}, \ \
r_{\pi}=.524~{\rm fm}.
\nonumber
\end{eqnarray}
One sees that agreement
with experimental data for the form factor and radius
$r_\pi^{\rm exp}=.656$~fm is quite satisfactory.

An improvement of the radius and form factor at small
$Q^2$ can come from the
diagram with intermediate $\rho$-meson (Fig.~7).
It is known~\cite{gross,efiv} that its contribution to $F_\pi(Q^2)$
can be important in the region
$Q^2<5{\rm Gev}^2$. An estimation of the diagram with $\rho$-meson
within the model under consideration also indicates
a diminishing of the form factor, which is maximal
(about $6\%$) in the region of $Q^2\approx 2 {\rm Gev}^2$.

Numerically
the contribution of
the bubble diagrams is
very small. It is maximal
at $Q^2\sim 2 {\rm Gev}^2$ and
is of order $10^{-3}$.
Thus within the model under consideration
the triangle diagram gives the main contribution
to the form factor for the values of $Q^2$ shown in Fig.~6.
One sees that the calculated form factor (the solid line)
smoothly approaches the experimental fit (dashed line) at large $Q^2$.
This behaviour seems unexpected. As is known from the studies,
based on the effective meson-quark quantum field
models~\cite{gross,valera,efiv,roberts},
the triangle diagram should decay
stronger
than
$1/Q^2$ --
the asymptotics of the experimental fit.
In our case, a naive estimation based on the
ultraviolet behaviour of the quark propagator~(\ref{qsol}) and vertex
(\ref{ffactor}) gives $(Q^2)^{-2}$.
However, the homogeneous vacuum gluon field changes the asymptotics of the
triangle diagram cardinally.

To demonstrate this let us consider the function $F_\pi^\triangle$
in the limit
$Q^2\gg \Lambda^2$. It is convenient to rewrite Eq.~(\ref{c3})
as
\begin{eqnarray}
\label{asimpt}
F_\pi^\triangle(Q^2)&=&\frac{h_\pi^2}{4\pi^2}
{\rm Tr}_v
\int\limits_{0}^{1}...\int\limits_{0}^{1}dt_1dt_2
ds_1ds_2ds_3
\left[\left(\frac{1-s_2}{1+s_2}\right)
\left(\frac{1-s_3}{1+s_3}\right)\right]^{m^2_u/4v}
\nonumber\\
&&\times
\Phi^\triangle(Q^2,M_\pi^2,m_u^2;s_i,t_i)
\exp\left\{-\frac{Q^2}{2v}\Omega(Q^2,s,t)\right\},
\\
\Phi^\triangle&=&
\exp\left(-Q^2\psi\right)
\left[Q^2\Phi_1(M^2_\pi,Q^2;s,t)
+m_u^2\Phi_2(M^2_\pi,Q^2;s,t)
\right.
\nonumber\\
&&\left.
+M_\pi^2\Phi_3(M^2_\pi,Q^2;s,t)
+\Phi_4(M^2_\pi,Q^2;s,t)\right].
\nonumber
\end{eqnarray}
Function $\Omega$ has the form
\begin{eqnarray}
\label{exp}
&&\Omega=2v(\varphi-\psi)
+\frac{m_u^2}{2Q^2}{\rm ln}\frac{1+s_1}{1-s_1}=
\frac{1-s_1}{A_1+s_1A_2}
+\frac{m_u^2}{2Q^2}\ln\frac{1+s_1}{1-s_1},
\end{eqnarray}
where $A_1$ and $A_2$ are functions of
$t_1$, $t_2$, $s_2$, and $s_3$ and do not
depend on $s_1$.
Function  $\phi$ is given in Eq.~(\ref{c3}), and
$$
\psi=[s_1s_2s_3+vt_1s_2(s_1+s_3)+vt_2s_3(s_1+s_2)
+v^2t_1t_2(s_1+s_2+s_3+s_1s_2s_3)]/2v\chi
$$
comes from the hyperbolic functions which have appeared in the integrand
due to averaging of the quark loop
over directions of the vacuum field.
For details we refer to Eqs.~(\ref{ave}) and (\ref{avec}) in
Appendices~B and C. We simply joined the exponentially increasing part
of these hyperbolic functions with the exponent in Eq.~(\ref{c3}).
It should be stressed that the origin of such an
exponentially increasing with
$Q^2\to\infty$ terms in the integrand is the presence of
phase factor $\exp(ix_\mu B_{\mu\nu}y_\nu)$ in the quark propagator
(\ref{qsol}) and covariant derivatives
$\nabla=\partial-iB$ in the vertices (\ref{ffactor}).
From the physical point of view, this simply means that in the presence
of external field the translation invariance holds for the gauge
invariant quantities but not for the
vertices and propagators separately. As is stressed in Appendix~C, this is
the reason why the energy-momentum is not conserved in the separate
vertices, but the conservation law is still valid for
a whole diagram, if it is gauge invariant.

One notices that the function $\Omega$ has a minimum at
$s_1=s_1^\ast$:
\begin{eqnarray}
&&\frac{\partial\Omega}{\partial s_1}
\Big\vert_{s_1=s_1^\ast}=0, \
s_1^\ast=1-\frac{m_u^2}{2Q^2}(A_1+A_2),
\nonumber\\
&&\frac{\partial^2\Omega}
{\partial s_1^2}\Big\vert_{s_1=s_1^\ast}=
\frac{2Q^2}{m_u^2(A_1+A_2)^2} > 0,
\nonumber\\
&&A_1+A_2=\frac{(1+s_2)(1+s_3)(1+2vt_1)(1+2vt_2)}
{s_2s_3+vt_1s_2(1-s_3)+vt_2s_3(1-s_2)+v^2t_1t_2(1-s_2)(1-s_3)},
\nonumber\\
&&\forall t_1,t_2,s_2,s_3,Q^2 \ s_1^\ast\in[0,1], \
\lim_{Q^2\to\infty}s_1^*=1,
\end{eqnarray}
which allows to integrate over variable $s_1$
using the saddle-point approximation.
The result is
\begin{eqnarray}
&&F_\pi^\triangle(Q^2)=
{\rm Tr}_v
\int \limits_{0}^{1}...\int \limits_{0}^{1}dt_1dt_2
ds_2ds_3\Phi^\triangle(Q^2,M_\pi^2,m_u^2;s_1^\ast,s_2,s_3,t_1,t_2)
\nonumber\\
&&\times
\left(\frac{m_u^2}{2Q^2}\right)^{m_u^2/4v}
\frac{\sqrt{2\pi vm_u^2(A_1+A_2)^2}}{Q^2}
\exp\left\{-\frac{m_u^2}{4v}
\left(1-{\rm ln}\frac{A_1+A_2}{2}\right)\right\}.
\nonumber
\end{eqnarray}
Since functions $\Phi_i$ contained in $\Phi^\triangle$
(see Eq.~(\ref{c3}) and Appendix~C) have the following
asymptotic form
\begin{eqnarray}
&&\Phi_2
\exp\left(-Q^2\psi\right)\Big|_{s_1=s_1^*}
\sim 1/Q^2(1-s_1^{\ast 2})=
2/m_u^2(A_1+A_2)={\rm const},
\nonumber\\
&&\Phi_{1,3}
\exp\left(-Q^2\psi\right)\Big|_{s_1=s_1^*}
\sim 1/Q^2, \
\Phi_4
\exp\left(-Q^2\psi\right)\Big|_{s_1=s_1^*}
\sim{\rm const},
\nonumber
\end{eqnarray}
hence  $\Phi^\triangle$ does not depend on $Q^2$ in the leading order
\begin{eqnarray}
&&\lim_{Q^2\to\infty}
\Phi^\triangle(Q^2,M_\pi^2,m_u^2;s_1^\ast,s_2,s_3,t_1,t_2)
=\Phi^\triangle_{\rm as}(M_\pi^2,m_u^2;s_2,s_3,t_1,t_2)+O(1/Q^2).
\nonumber
\end{eqnarray}
Finally,
the asymptotic formula for the triangle diagram reads
\begin{eqnarray}
\label{asimptf}
&&F_\pi^\triangle(Q^2/\Lambda^2)=
{\rm Tr}_v \frac{C^\triangle(M^2_\pi/\Lambda^2,m_u^2/\Lambda^2)}
{\left(Q^2/\Lambda^2\right)^{1+m_u^2/4v\Lambda^2}},
\end{eqnarray}
where factor $C^\triangle$ is independent of $Q^2$,
\begin{eqnarray}
&&C^\triangle=h_\pi^2
\frac{\sqrt{v}}{\sqrt{2\pi}\pi}
\left(\frac{m_u}{2\Lambda}\right)^{1+m_u^2/2v\Lambda^2}
e^{-m_u^2/4v\Lambda^2}
\nonumber\\
&&\times\int\limits_{0}^{1}...\int\limits_{0}^{1}dt_1dt_2
ds_2ds_3\Phi_{\rm as}(M_\pi^2/\Lambda^2,m_u^2/\Lambda^2;s_2,s_3,t_1,t_2)
(A_1+A_2)^{1+m_u^2/4v\Lambda^2}.
\nonumber
\end{eqnarray}
Here we have returned to the dimensionful notation for the masses
and momentum $Q$.
For the parameter values from Table~I
we obtain the following result
\begin{eqnarray}
&&F_\pi^\triangle(Q^2/\Lambda^2)=
\frac{2.96}{\left(Q^2/\Lambda^2\right)^{1.1435}}.
\end{eqnarray}
This asymptotic formula fits well the solid curve in Fig.~6
for $Q^2> 5{\rm Gev}^2$.

Thus the main effect of the vacuum field under consideration
in the charge form factor is
an increasing of the contribution of triangle diagram to the form
factor at large $Q^2$. As has been explained,
this is clearly due to
the presence of the background field both in the quark propagators and
nonlocal meson-quark vertices, which causes
the specific interplay of
translation and color gauge invariance in the quark loops.
A comparison with the large $Q^2$ behaviour of the $\pi\gamma\gamma$
triangle diagram, which is usual $1/Q^2$ (see the previous section),
indicates that a number of nonlocal vertices in a loop is of crucial
importance.

The contribution of bubble graph (see Fig.~5) to asymptotic
behaviour is easy to derive if to rewrite Eq.~(\ref{b}) in the form
\begin{eqnarray}
&&F_\pi^\circ(Q^2)={\rm Tr}_v
\int\limits_{0}^{1}...\int\limits_{0}^{1}dt_1dt_2
ds_1ds_2d\beta \ \Phi^\circ(Q^2\beta,M_\pi^2,m_u^2;s_i,t_i)
\exp\left\{-Q^2\varphi^\circ(\beta,s_i,t_i)\right\},
\nonumber
\end{eqnarray}
which underlines that the preexponential factor
$\Phi^\circ$ depends on $Q^2$
in the combination $Q^2\beta$, as is seen from
functions $\Phi_i^\circ$ (see Eq.~(\ref{b}) and  Appendix~C).
Here $\beta$ is a proper time corresponding to the vertex operator
$\Gamma_\mu$ given by Eq.~(\ref{finalb}).
The smallest value of the function $\varphi^\circ$
corresponds to the point $\beta=0$ for any $t_1$, $t_2$, $s_1$, and $s_2$:
\begin{eqnarray}
\varphi^\circ(\beta,t_1,t_2,s_1,s_2)&=&\beta
\tilde\varphi^\circ(\beta,t_1,t_2,s_1,s_2),
\nonumber \\
\frac{\partial\varphi^\circ}{\partial\beta}
\Big\vert_{\beta=0}&=&\tilde\varphi^\circ(0,t_1,t_2,s_1,s_2)
\nonumber\\
&=&\frac{vt_2}{2v\chi^{\circ}}\left[s_1+v(t_1+t_2)(1+s_1s_2)
+2v^2t_1t_2(s_1+s_2)\right]>0.
\nonumber
\end{eqnarray}
Therefore the leading term takes the form
\begin{eqnarray}
\label{asb}
&&F_\pi^\circ(Q^2/\Lambda^2)=
C^\circ(M^2_\pi/\Lambda^2,m_u^2/\Lambda^2)
\frac{\Lambda^2}{Q^2}+O\left(\left(\frac{\Lambda^2}{Q^2}\right)^2\right),
\nonumber\\
&&C^\circ={\rm Tr}_v
\int\limits_{0}^{1}...\int\limits_{0}^{1}dt_1dt_2ds_1ds_2
\int\limits_{0}^{\infty}d\beta
\Phi^\circ(\beta,M^2_\pi/\Lambda^2,m_u^2/\Lambda^2;
t_1,t_2,s_1,s_2)
\nonumber \\
&&\times\exp\left[-\beta\tilde\varphi^\circ(0,t_1,t_2,s_1,s_2)\right]
\approx 0.3.
\nonumber
\end{eqnarray}
We conclude that the absolute asymptotics of the charge form factor
is defined by the bubble diagram.
The limit $Q^2\gg\Lambda^2$ of this diagram
is due to
the ultraviolet regime ($\beta\to 0$) in the vertex $\Gamma_\mu$.
The vertex is determined directly by the gluon propagator,
and the $1/Q^2$ dependence appears as a manifestation of the
ultraviolet behavior of
the gluon propagator.  This is in agreement with the mechanism of hard
rescattering and quark
counting rules~\cite{matv}.
Namely, the asymptotic behaviour of the form factor is
determined by the one-gluon exchange between quarks inside
a pion. However, in the experimentally observed region
the triangle diagram dominates in the
form factor.\\

\newpage

\noindent
{\Large\bf Acknowledgments}\\

The authors would like to thank V.~Lyubovitskij and S.~Mikhailov
for numerous illuminating discussions.
We are also grateful to A.~Bakulev, A.~Dorokhov, A.~Efremov,
M.~Ivanov, N.~Kochelev, P.~Minkowski, R.~Ruskov
 and S.~Solunin for useful comments. J.V.B. thanks I.~Anikin for discussions.
This work was supported in part
by the RFBR grant No.~96-02-17435-a.

\section{Appendix A}

Below we present a procedure for decomposition
of the vertices $V_{aJ\ell n}$ written in Eq.~(\ref{v-decomp}).
We will consider the vertex for pion with $\ell=n=0$ and
$\xi_f=\xi_{f^\prime}=1/2$. An
extension to the general case is straightforward.
As is seen from
Eqs.~(\ref{i1},\ref{ffactor}) we have to decompose
the operator $F_{00}$ in a series over
electromagnetic field $A$. The presence of the background gluon
field in the vertex is irrelevant to the question under consideration.
For simplicity we drop the gluon field below. At the end one have to
change the partial derivatives to the covariant ones.

The most convenient way is to use
the path ordered exponential representation
($e_1, e_2$ are the charges of $u$ and $d$ quarks):
\begin{eqnarray}
\label{pathod1}
&&F_{00}
\left[\left(\stackrel{\leftrightarrow}{\partial}_\mu(x)
+i(e_1+e_2)A_\mu(x)\right)^2/4\Lambda^2\right]
\nonumber\\
&&=\int\limits_{0}^{1}dt
P_\beta
\exp\left[\frac{t}{4\Lambda^2}
\int\limits_0^1d\beta
\left(\stackrel{\leftrightarrow}{\partial}(x(\beta))
+i(e_1+e_2)A(x(\beta))\right)^2\right]
\nonumber \\
&&=P_\beta\int\limits_{0}^{1}dt\int\frac{d^4a}{\pi^2}
e^{-\int\limits_{0}^{1}d\beta a^2(\beta)}
e^{\frac{\sqrt t}{\Lambda}
\int\limits_{0}^{1}d\beta a(\beta)
\left(\stackrel{\leftrightarrow}{\partial}
+i(e_1+e_2)A\right)}
\nonumber \\
&&=P_\beta\int\limits_{0}^{1}dt\int d\sigma_a
e^{\frac{\sqrt t}{\Lambda}
\int\limits_{0}^{1}d\beta a(\beta)
\left(\stackrel{\leftarrow}{\partial}
+ie_1A\right)}
e^{-\frac{\sqrt t}{\Lambda}
\int\limits_{0}^{1}d\beta a(\beta)
\left(\stackrel{\rightarrow}{\partial}
-ie_2A\right)}
\nonumber
\end{eqnarray}
Using this representation, one gets a nonambiguous decomposition for
 matrix element of the vertex operator:
\begin{eqnarray}
\label{vb}
&&e^{-ipx}F_{00}
\left[\left(\stackrel{\leftrightarrow}{\partial}_\mu(x)
+i(e_1+e_2)A_\mu(x)\right)^2/4\Lambda^2\right]e^{ikx}
\nonumber\\
&&=\int\limits_{0}^{1}dt\int d\sigma_a
\exp\left[-ip\left(x+\frac{\sqrt t}{\Lambda}
\int\limits_{0}^{1}d\beta a(\beta)\right)
+ik\left(x-\frac{\sqrt t}{\Lambda}\int\limits_{0}^{1}d\beta a(\beta)\right)
\right.
\nonumber \\
&&
\ \ \ \ \ \ \ \ \ \ \ \ \ \ \ \ \ \ \ \ \ \ \ \
+i\frac{\sqrt t}{\Lambda}e_1\int\limits_{0}^{1}d\beta a(\beta)
A\left(x+\frac{\sqrt t}{\Lambda}\int\limits_{0}^{\beta}d\beta'
a(\beta')\right)
\nonumber \\
&&
\ \ \ \ \ \ \ \ \ \ \ \ \ \ \ \ \ \ \ \ \ \ \ \
\left.
+i\frac{\sqrt t}{\Lambda}e_2\int\limits_{0}^{1}d\beta
a(\beta)A\left(x-\frac{\sqrt t}{\Lambda}
\int\limits_{0}^{\beta}d\beta'a(\beta')\right)\right]
\nonumber \\
&&
=\int\limits_{0}^{1}dt\int d\sigma_a
\exp\left\{-i(p-k)x-i\frac{\sqrt t}{\Lambda}(p+k)
\int\limits_{0}^{1}d\beta a(\beta)\right\}
\nonumber \\
&&
\times\left[1+
i\frac{\sqrt t}{\Lambda}e_1\int\limits_{0}^{1}d\beta a(\beta)
A\left(x+\frac{\sqrt t}{\Lambda}\int\limits_{0}^{\beta}d\beta'
a(\beta')\right)
\right.
\nonumber \\
&&
\left.
+i\frac{\sqrt t}{\Lambda}e_2\int\limits_{0}^{1}d\beta
a(\beta)A\left(x-\frac{\sqrt t}{\Lambda}
\int\limits_{0}^{\beta}d\beta'a(\beta')\right)+O(e_1^2)+O(e_2^2)\right].
\end{eqnarray}
If we now introduce Fourier transformed electromagnetic field,
$$
A_\mu(y)=\int\frac{d^4q}{(2\pi)^4}{\rm e}^{iqy}\tilde A_\mu(q),
$$
then a calculation of coefficients of decomposition to any order
in $A$ is reduced to evaluation of Gaussian path integrals.
These integrals can be performed expanding the paths
$a_\mu(\beta)$ into trigonometric series
$$
a_\mu(\beta)=a^0_\mu+\sum_{n=0}^{\infty}
[u^n_\mu cos 2\pi n\beta+v^n_\mu sin 2\pi n\beta]=a^0_\mu+\bar a_\mu(\beta).
$$
In the linear order in $A$ one has to evaluate the integrals:
\begin{eqnarray}
&& \int\limits_{0}^{1}dt\int d\sigma_a
\exp\left[-i(p-k)x-i\frac{\sqrt t}{\Lambda}a_0(p+k)\right]
=\int\limits_{0}^{1}dt\exp\left[-i(p-k)x-\frac{t}{4\Lambda^2}
(p+k)^2\right]
\nonumber \\
&&=e^{-ipx}F_{00}(-(p+k)^2)e^{ikx}
=e^{-ipx}F_{00}\left(\stackrel{\leftrightarrow}{\partial^2}(x)
\right)e^{ikx},
\nonumber \\
&&\int d\sigma_a a^0_\mu
\exp\left[-i\frac{\sqrt t}{\Lambda}a_0(p+k-\beta q)\right]
\nonumber \\
&&=-i\frac{\sqrt t}{2\Lambda}(p+k-\beta q)_\mu \
\exp\left[-\frac{t}{4\Lambda^2}(p+k-\beta q)^2\right],
\nonumber \\
&&\int d\sigma_a
\exp\left[-i\frac{\sqrt t}{\Lambda}a_0(p+k)+
i\frac{\sqrt t}{\Lambda}\beta a^0 q\right]
=\exp\left[-\frac{t}{4\Lambda^2}(p+k-\beta q)^2\right],
\nonumber \\
&&\int d\sigma_{\bar a}
\exp\left[iq\frac{\sqrt t}{\Lambda}\int\limits_{0}^{\beta}d\beta'
\bar a(\beta')\right]=\exp\left[-\frac{t}{4\Lambda^2}
\beta(1-\beta)q^2\right],
\nonumber \\
&&\int d\sigma_{\bar a} \bar a_\mu
\exp\left[iq\frac{\sqrt t}{\Lambda}\int\limits_{0}^{\beta}d\beta'
\bar a(\beta')\right]=i\frac{\sqrt t}{4\Lambda}(1-2\beta)q_\mu
\exp\left[-\frac{t}{4\Lambda^2}\beta(1-\beta)q^2\right],
\nonumber
\end{eqnarray}
so that
\begin{eqnarray}
\label{sigma}
&&\int d\sigma_a a_\mu(\beta)
\exp\left[-i\frac{\sqrt t}{\Lambda}(p+k)\int_{0}^{1}d\beta a(\beta)
+iq\frac{\sqrt t}{\Lambda}\int\limits_{0}^{\beta}d\beta'
a(\beta')\right]=
\nonumber \\
&&-i\frac{\sqrt t}{4\Lambda}(2(p+k)_\mu-q_\mu) \
\exp\left[-\frac{t}{4\Lambda^2}[(p+k)^2-\beta(2(p+k)-q)q]\right].
\end{eqnarray}
Inserting Eq.~(\ref{sigma}) into Eq.~(\ref{vb}) one arrives at
the representation
\begin{eqnarray}
\label{bb}
&&e^{-ipx}F_{00}
\left(\left[\stackrel{\leftrightarrow}{\partial}
+i(e_1+e_2)A\right]^2\right)e^{ikx}
=e^{-ipx}F_{00}(-(p+k)^2)e^{ikx}
\nonumber \\
&&+(e_1-e_2)\int\frac{d^4q}{(2\pi)^4}e^{ix(q+k-p)}\tilde A_\mu(q)
\int\limits_0^1\int\limits_0^1d\beta dt
\frac{t}{4\Lambda^2}[2(p+k)_\mu-q_\mu]
\nonumber \\
&&\times\exp\left[-\frac{t}{4\Lambda^2}[(p+k)^2-\beta(2(p+k)-q)]q\right]
+O(e_1^2)+O(e_2^2)=
\nonumber \\
&&e^{-ipx}\left[F_{00}[\stackrel{\leftrightarrow}{\partial}^2(x)]
+(e_1-e_2)\int d^4y A_\mu(y)\Gamma_\mu(x,y)\right]e^{ikx}
+O(e_1^2)+O(e_2^2)
\nonumber
\end{eqnarray}
with
\begin{eqnarray}
&&\Gamma_\mu(x,y)=
\int\frac{d^4q}{(2\pi)^4}e^{iq(x-y)}
\int\limits_0^1\int\limits_0^1dt d\beta
\frac{t}{4\Lambda^2}[2i\stackrel{\leftrightarrow}{\partial}_\mu(x)-q_\mu]
\nonumber \\
&& \ \ \ \ \ \ \times\exp\left[\frac{t}{4\Lambda^2}
\left(\stackrel{\leftrightarrow}{\partial}^2(x)
+\beta (2i\stackrel{\leftrightarrow}{\partial}(x)-q)q\right)\right].
\end{eqnarray}
To get Eq.~(\ref{finalb}) one have to change partial derivatives to
covariant derivatives $\nabla$ with the vacuum gluon field.

Equation (\ref{bb}) can be written in the form
\begin{eqnarray}
\label{bubble}
&&e^{-ipx}\left\{F_{00}(-(p+k)^2)
+(e_1-e_2)\int\frac{d^4k}{(2\pi)^4}
e^{iqx}\tilde A_\mu(q)\frac{2(p+k)_\mu-q_\mu}{2(p+k)q-q^2}
\right.
\nonumber \\
&&
\left.
\times\left[F_{00}(-(p+k-q)^2)-F_{00}(-(p+k)^2)\right]\right\}
e^{ikx}.
\end{eqnarray}
This formula for vertex function
coincides with the representations obtained earlier in
papers~\cite{gross,valera}.
However, unlike the algorithms used in~\cite{gross,valera,ter}
above procedure can be applied regularly to calculate  the higher
order terms, i.e., vertices with two, three or more
photons. The calculation is reduced to
evaluation of Gaussian integrals.

\section{Appendix B}
Here we give some details of calculation of the diagram in Fig~3.
The contribution of the triangle diagram
to the $\pi\gamma\gamma$ vertex can be represented in
the form (below we use the dimensionless, i.e. divided by
$\Lambda$, momenta and masses)
\begin{eqnarray}
\label{T}
&&T_\pi^{\mu\nu}(p,k_1,k_2)=
\sqrt{2}h_\pi\Lambda
\int d\sigma_{B}\int\frac{d^4p_1d^4p_2d^4p_3}{(2\pi)^{12}}
\tilde R(k_1,k_2,p;p_1,p_2,p_3)
\nonumber \\
&&
\times{\rm Tr}\tau_3 i\gamma_5\tilde H(p_1)
{\cal Q}\gamma_\mu\tilde H(p_2)
{\cal Q}\gamma_\nu\tilde H(p_3).
\end{eqnarray}
\begin{eqnarray}
\label{Re}
&&\tilde R=\int d^4xd^4yd^4z e^{i(px+k_1z-k_2y)}
R(x,y,z;p_1,p_2,p_3), \\
\label{RRe}
&&R(x,y,z;p_1,p_2,p_3)=\exp\left[iz\hat Bx+ip_3(z-x)\right] F_{00}
\left(\frac{\stackrel{\leftrightarrow}{\nabla}^2(x)}{\Lambda^2}\right)
\exp\left[ix\hat By+ip_1(x-y)\right] \nonumber\\
&&\times\exp\left[iy\hat Bz+ip_2(y-z)\right],
\end{eqnarray}
where ${\cal Q}={\rm diag}(2/3,-1/3)$ is the charge matrix,
quark propagator $\tilde H$
is defined by Eq.~(\ref{qsol}), and $F_{00}$
is the vertex operator (\ref{ffactor}).

Acting by $F_{00}$ onto the
exponents in Eq.~(\ref{RRe}) and integrating over
the space-time coordinates in Eq.~(\ref{Re}), we get
\begin{eqnarray}
\label{RRRe}
&&\tilde R=
\delta^{(4)}(p+k_1+k_2)
\frac{16\pi^4}{v^4}\int\limits_{0}^{1}dt
\exp\left[-\frac{1}{4}\beta^T{\cal M}_1 \beta
-\frac{t}{4}(p_1+p_3)^2\right],
\end{eqnarray}
where $\beta^T=(\beta_1^T \beta_2^T)$,
\begin{eqnarray}
&&\beta_{1\mu}=(k_2+p_2-p_1)_\mu-ivtf_\mu(p_1+p_3)/2,
\nonumber\\
&&\beta_{2\mu}=(k_1+p_2-p_3)_\mu+ivtf_{\mu}(p_1+p_3)/2,
\nonumber\\
&&{\cal M}_1=\frac{1}{v}\left(
\begin{array}{cc}
vtI & M_1 \\
M^T_1 & vtI
\end{array} \right), \
M_1^{\mu\nu}=vt\delta^{\mu\nu}-2if^{\mu\nu}.
\nonumber
\end{eqnarray}
Inserting Eq.~(\ref{RRRe}) into Eq.~(\ref{T}), one obtains
\begin{eqnarray}
\label{TT}
&&T_\pi^{\mu\nu}(p,k_1,k_2)=
\delta^{(4)}(p+k_1+k_2){\rm Tr}_v
\frac{\sqrt 2}{v^4}h_\pi
\int d\sigma_B\int\limits_{0}^{1}dt
\int\frac{d^4p_1d^4p_2d^4p_3}{(2\pi)^{12}}
\nonumber \\
&&\times
\exp\left[-\frac{1}{4}\beta^T{\cal M}_1 \beta
-\frac{t}{4}(p_1+p_3)^2\right]
{\rm Tr}\tau_3 i\gamma_5\tilde H(p_1)
{\cal Q}\gamma_\mu\tilde H(p_2)
{\cal Q}\gamma_\nu\tilde H(p_3).
\nonumber \\
\end{eqnarray}
A calculation of traces
gives
\begin{eqnarray}
\label{TTT}
&&T_\pi^{\mu\nu}(p,k_1,k_2)=i \epsilon_{\mu\nu\alpha\beta}
\delta^{(4)}(k_1+k_2+p)
\nonumber \\
&&\times{\rm Tr}_v\frac{\Lambda h_\pi}{3\sqrt 2 v^7 }
{\rm Tr}_v\int d\sigma_B
\int \limits_{0}^{1}...\int \limits_{0}^{1}dtds_1ds_2ds_3
\left[\left(\frac{1-s_1}{1+s_1}\right)
\left(\frac{1-s_2}{1+s_2}\right)
\left(\frac{1-s_3}{1+s_3}\right)\right]^{m^2_u/4v}
\nonumber\\
&&\times \int\frac{d^4p_1d^4p_2d^4p_3}{(2\pi)^8}
{\cal W}_{\alpha\beta}(p_1,p_2,p_3;m_u)
\exp\left[-P^T{\cal M}_2 P+
P^TY-h(p,k_1)\right],
\end{eqnarray}
where
\begin{eqnarray}
\label{W}
&&{\cal W}^{\alpha\beta}=
\frac{m_u}{1-s_1^2}p_2^\alpha p_3^\beta-
\frac{m_u}{1-s_2^2}p_1^\alpha p_3^\beta+
\frac{m_u}{1-s_3^2}p_1^\alpha p_2^\beta,
\nonumber\\
&&P^T=\left(
p^T_1 p_2^T p_3^T \right), \
Y^T=\left(
Y^T_1 Y^T_2 Y^T_3 \right),
\nonumber\\
&&Y_{1\mu}=t(p+2k_1)_\mu-if_\mu(k_1)/v,
Y_{2\mu}=-t(p+2k_1)_\mu+if_\mu(p)/v,
\nonumber\\
&&Y_{3\mu}=t(p+2k_1)_\mu-if_\mu(p-k_1)/v,
h=tp^2/4+tk_1^2+tpk_1+ik_{1\mu}f_{\mu\nu}p_\nu/v,
\nonumber\\
&&{\cal M}_2=\frac{1}{2v}\left(
\begin{array}{ccc}
(2vt+s_1) \ I & -M_2 & M_2 \\
-M^T_2 & (2vt+s_2) \ I & -M_2 \\
M^T_2 & -M^T_2 & (2vt+s_3) \ I
\end{array} \right),
M^{\mu\nu}_2=2vt\delta^{\mu\nu}-if^{\mu\nu}.
\nonumber
\end{eqnarray}
Now one has to integrate over $p_1$, $p_2$,
$p_3$, and average over directions of the vacuum field
using the following formulas
\begin{eqnarray}
\label{aveb}
&&\int d\sigma_B \exp(2is_2\psi p_\gamma f_{\gamma\sigma}k_{1\sigma})=
{\cal F}_0(s_2\psi;k_1^2,M_\pi),
\nonumber \\
&& \int d\sigma_B \epsilon^{\mu\nu\alpha\beta}
if^{\alpha\beta}\exp(2is_2\psi p_\gamma f_{\gamma\sigma}k_{1\sigma})=
-4\epsilon^{\mu\nu\alpha\beta}k_1^\alpha k_2^\beta
{\cal F}_1(s_2\psi;k_1^2,M_\pi),
\nonumber \\
&&\int d\sigma_B \epsilon^{\mu\nu\alpha\beta}k_1^\alpha p^\rho
if^{\beta\rho}\exp(2is_2\psi p_\gamma f_{\gamma\sigma}k_{1\sigma})=
2\epsilon^{\mu\nu\alpha\beta}k_1^\alpha k_2^\beta(k_1p)
{\cal F}_1(s_2\psi;k_1^2,M_\pi),
\nonumber \\
&&\int d\sigma_B \epsilon^{\mu\nu\alpha\beta}p^\rho k_1^\xi
f^{\alpha\rho}f^{\beta\xi}
\exp(2is_2\psi p_\gamma f_{\gamma\sigma}k_{1\sigma})=
\epsilon^{\mu\nu\alpha\beta}k_1^\alpha k_2^\beta
{\cal F}_2(s_2\psi;k_1^2,M_\pi),
\nonumber \\
&&{\cal F}_0=\frac{1}{s_2\psi(k_1^2+M^2_\pi)}
\sinh\left[s_2\psi(k_1^2+M^2_\pi)\right],
\nonumber \\
&&{\cal F}_1=\frac{1}{s_2\psi(k_1^2+M^2_\pi)^2}
\left[\cosh\left[s_2\psi(k_1^2+M^2_\pi)\right]-{\cal F}_0\right],
\nonumber \\
&&{\cal F}_2={\cal F}_0+4{\cal F}_1/s_2\psi,
\nonumber \\
&&\psi=(s_1s_3+vt(s_1+s_3))/2v\chi, \
\chi=s_1+s_2+s_3+s_1s_2s_3+2vt(1+s_1s_2+s_1s_3+s_2s_3).
\nonumber
\end{eqnarray}
These equations can be derived
from the generating formula
\begin{equation}
\label{ave}
\frac{1}{4\pi}\int_{0}^{2\pi}d\varphi\int_{0}^{\pi}d\theta\sin\theta
\exp\left(if_{\mu\nu}J_{\mu\nu}\right)=\frac{\sin
\sqrt{2\left(J_{\mu\nu}J_{\mu\nu}\pm\tilde J_{\mu\nu}J_{\mu\nu}\right)}}
{\sqrt{2\left(J_{\mu\nu}J_{\mu\nu}\pm\tilde J_{\mu\nu}J_{\mu\nu}\right)}},
\end{equation}
where $\varphi$ and $\theta$ are the spherical angles, $J_{\mu\nu}$ is an
anti-symmetric tensor, $\tilde J_{\mu\nu}$ is a dual tensor and
$\pm$ corresponds to the self- and anti-self-dual vacuum field.

After these manipulations we arrive
at the expressions for $T_\pi^{\mu\nu}(k_1,k_2,p)$,  Eq.~(\ref{1}),
and for the transition form factor $F_{\gamma\pi}(Q^2)$, Eq.~(\ref{5}),
with functions $\Phi_i$ ($i=1,2,3$) defined by the relations
\begin{eqnarray}
&&\Phi_i(M_\pi^2,Q^2;s_1,s_2,s_3,t)=\left[{\cal F}_0P_i
+{\cal F}_1R_i
+{\cal F}_2L_i\right]/6v\chi^4.
\nonumber
\end{eqnarray}
The polynomials $P_i,R_i,L_i$ have the form
\begin{eqnarray}
&&P_1=vts_1^2s_2+(s_1+vt)[s_1+s_2+s_3+2vt(1+s_1s_3)],
\nonumber \\
&&P_2=s_2[(1+vt(s_1+s_3))(s_2+2vt)+s_1(1+2vts_3)+s_3(1+2vts_1)],
\nonumber \\
&&P_3=vts_2s_3^2+(s_3+vt)[s_1+s_2+s_3+2vt(1+s_1s_3)],
\nonumber \\
&&R_i=M^2_\pi Q_{1i}(s,t)-Q^2 Q_{2i}(s,t)+Q_{3i}(s,t),
\nonumber \\
&&Q_{11}=(s_1+2vt)[2s_1s_3+vt(s_1+s_3-s_2(1+s_1s_3))],
\nonumber \\
&&Q_{12}=2s_1s_2s_3-2vt[2s_1s_3+s_2^2(1+s_1s_3)]-2v^2t^2(1+s_2^2)(s_1+s_3),
\nonumber \\
&&Q_{13}=(s_3+2vt)[2s_1s_3+vt(s_1+s_3-s_2(1+s_1s_3))],
\nonumber \\
&&Q_{21}=s_2(s_1+2vt)[s_3+vt(1+s_1s_3)]+s_2(1+2vts_1)[s_1s_3+vt(s_1+s_3)]
\nonumber \\
&&+vt(s_1+s_2)[s_1+s_3+2vt(1+s_1s_3)],
\nonumber \\
&&Q_{22}=2s_2^2s_3+2vts_2[s_1-s_3+2s_2(1+s_1s_3)]+4v^2t^2s_2^2(s_1+s_3),
\nonumber \\
&&Q_{23}=[s_1+s_3+2vt(1+s_1s_3)][s_2s_3+vt(s_2-s_3)]
+s_2(s_3+2vt)[s_3+vt(1+s_1s_3)]
\nonumber \\
&&-s_2(1+2vts_3)[s_1s_3+vt(s_1+s_3)],
\nonumber \\
&&Q_{31}=4v\chi(s_1+2vt),
\nonumber \\
&&Q_{32}=4v\chi(s_2-2vt),
\nonumber \\
&&Q_{33}=4v\chi(s_3+2vt),
\nonumber \\
&&L_1=s_2(s_1+2vt)[s_1s_3+vt(s_1+s_3)],
\nonumber \\
&&L_2=s_2(s_2-2vt)[s_1s_3+vt(s_1+s_3)],
\nonumber \\
&&L_3=s_2(s_3+2vt)[s_1s_3+vt(s_1+s_3)].
\nonumber
\end{eqnarray}
The polynomial $\chi$ is written in Eq.~(\ref{5}).

\section{Appendix C}

As it could be seen above, the background field complicates
calculation of diagrams. The reason is that
in the presence of external field the translation invariance is held only
for gauge invariant
quantities
like the diagrams in Fig.~5.
This is not true for propagators or vertices separately.
This means that the energy-momentum is concerved for the total
diagram but not for the separate vertices and the usual Feynman rules are
not valid. So that we have to begin calculation in the configuration space.
The contribution of the triangle diagram to the $\pi^+\pi^-\gamma$
vertex has the form
\begin{eqnarray}
\Lambda^\triangle_\mu(x,y,z)=h_\pi^2{\rm Tr}\int d\sigma_B
S(z,x)i\gamma_5 F_{00}\left(
\frac{\stackrel{\leftrightarrow}{\nabla}^2(x)}{\Lambda^2}\right)
S(x,y)i\gamma_5 F_{00}\left(
\frac{\stackrel{\leftrightarrow}{\nabla}^2(y)}{\Lambda^2}\right)
S(y,z)\gamma_\mu.
\nonumber
\end{eqnarray}
Fourier transformed expression reads
\begin{eqnarray}
\label{L}
\Lambda_\mu^\triangle(k_1,k_2,q)=
h^2_\pi{\rm Tr}[\tau^+,\tau^-]_{-}{\cal Q}
\int d\sigma_{B}\int\frac{d^4p_1d^4p_2d^4p_3}{(2\pi)^{12}}
\tilde R(k_1,k_2,q;p_1,p_2,p_3)
\nonumber \\
\times
i\gamma_5\tilde H(p_1)i\gamma_5\tilde H(p_2)
\gamma_\mu\tilde H(p_3), \\
\tilde R=\int d^4xd^4yd^4z e^{i(k_1x-k_2y+qz)}
R(x,y,z;p_1,p_2,p_3),
\nonumber\\
\label{RR}
R=\exp\left[iz\hat Bx+ip_3(z-x)\right] F_{00}
\left(\frac{\stackrel{\leftrightarrow}{\nabla}^2(x)}{\Lambda^2}\right)
\exp\left[ix\hat By+ip_1(x-y)\right] \nonumber\\
\times F_{00}
\left(\frac{\stackrel{\leftrightarrow}{\nabla}^2(y)}{\Lambda^2}\right)
\exp\left[iy\hat Bz+ip_2(y-z)\right],
\end{eqnarray}
where
${\cal Q}$ is the quark charge matrix, $\tilde H(p)$
is the quark propagator in the vacuum
field
(\ref{qsol}), and $F_{00}$
is the vertex operator (\ref{ffactor}).

The result of action of the operators $F_{00}$ and subsequent integration
over $x$, $y$ and $z$ looks as
\begin{eqnarray}
\label{RRR}
\tilde R=\delta^{(4)}(k_1-k_2+q)
\frac{16\pi^4}{v^4}
\int\limits_{0}^{1}\!\! \int\limits_{0}^{1}
\frac{dt_1dt_2}{(1+4v^2t_1t_2)^2}
\exp\left[-\frac{1}{4}\beta^T{\cal M}_1 \beta
-g(t_1,t_2;p_1,p_2,p_3)\right],
\end{eqnarray}
where
$\beta^{\rm T}=(\beta_1^{\rm T} \beta_2^{\rm T})$,
\begin{eqnarray}
&&{\cal M}_1=\frac{1}{v(1+4v^2t_1t_2)}\left(
\begin{array}{cc}
v(t_1+t_2) I & M_1 \\
M^T_1 & v(t_1+4t_2) \ I
\end{array} \right),
\nonumber \\
&&M_1^{\mu\nu}=v(t_1-2t_2)\delta^{\mu\nu}-2if^{\mu\nu},
\nonumber\\
&&\beta_{1\mu}=
[4(k_2+p_2-p_1)_\mu+v^2t_1t_2(k_2+2p_1+2p_2+2p_3)_\mu
\nonumber \\
&&-2ivt_1f_\mu(p_1+p_3)+4ivt_2f_\mu(p_1+p_2)]/
[4+v^2t_1t_2],
\nonumber \\
&&\beta_{2\mu}= [4(q+p_2-p_3)_\mu+v^2t_1t_2q_\mu
+2ivt_1f_\mu(p_1+p_3)+2ivt_2f_\mu(p_1+p_2)]
/[4+v^2t_1t_2],
\nonumber \\
&&g=[t_1(p_1+p_3)^2+t_2(p_1+p_2)^2
+ivt_1t_2(p_1+p_2)_\mu f_{\mu\nu}(p_1+p_3)_\nu]/[4+v^2t_1t_2].
\nonumber
\end{eqnarray}
Using Eq.~(\ref{RRR}) in Eq.~(\ref{L}) we arrive at
\begin{eqnarray}
\label{LL}
&&\Lambda_\mu^{\triangle}(k_1,k_2,q)=
\delta^{(4)}(k_1+q-k_2){\rm Tr}_v\frac{h_\pi^2}{v^4}
\int\limits_{0}^{1}\int\limits_{0}^{1}
\frac{dt_1dt_2}{(1+4v^2t_1t_2)^2}
\int d\sigma_{B}\int\frac{d^4p_1d^4p_2d^4p_3}{(2\pi)^{8}}
\nonumber\\
&&\times{\rm Tr}
i\gamma_5\tilde H(p_1)
i\gamma_5\tilde H(p_2)\gamma_\mu\tilde H(p_3)
\exp\left[-\frac{1}{4}\beta^{\rm T}{\cal M}_1\beta
-g(t_1,t_2;p_1,p_2,p_3)\right].
\end{eqnarray}
As has been discussed above, the delta function in front of the
integral in Eq.~(\ref{LL}) indicates the general conservation
of energy-momentum for the total diagram.
However,
we have to evaluate three momentum integrals instead of one
loop integral
in the usual case of zero external field. These Gaussian integrals
can be performed as follows.

Using representation~(\ref{qsol}) and calculating
the trace of  Dirac matrices,
one gets
\begin{eqnarray}
\label{ll}
&&\Lambda_\mu^{\triangle}(k_1,k_2,q)={\rm Tr}_v
\frac{h_\pi^2\Lambda}{2v^7}
\int d\sigma_{B}\int\limits_{0}^{1}\dots\int \limits_{0}^{1}
\frac{dt_1dt_2ds_1ds_2ds_3}{(1+4v^2t_1t_2)^2}
\nonumber\\
&&\times\left[\left(\frac{1-s_1}{1+s_1}\right)
\left(\frac{1-s_2}{1+s_2}\right)
\left(\frac{1-s_3}{1+s_3}\right)\right]^{m^2_u/4v}
\nonumber\\
&&\times\int\frac{d^4p_1d^4p_2d^4p_3}{(2\pi)^8}
{\cal W}_\mu(p_1,p_2,p_3)
\exp\left[-P^{\rm T}{\cal M}_2 P+
P^{\rm T}Y-h(k_1,q)\right],
\end{eqnarray}
where
$P^{\rm T}=\left(p^{\rm T}_1 p^{\rm T}_2 p^{\rm T}_3\right)$,
$Y^{\rm T}=\left(Y^{\rm T}_1 Y^{\rm T}_2 Y^{\rm T}_3 \right)$,
\begin{eqnarray}
&&Y^\mu_1=[v(t_1+t_2)k^\mu_1+v(2t_1-t_2)q^\mu
-i(1-2v^2t_1t_2)f^\mu(q)]/[v(1+4v^2t_1t_2)],
\nonumber \\
&&Y^\mu_2=-[v(t_1-t_2)k^\mu_1+v(2t_1+t_2)q^\mu
-2iv^2t_1t_2f^\mu(q)-if^\mu(k_1)]/[v(1+4v^2t_1t_2)],
\nonumber \\
&&Y^\mu_3=[v(t_1-t_2)k^\mu_1+v(2t_1+t_2)q^\mu
+i(1+2v^2t_1t_2)f^\mu(q)-if^\mu(k_1)]/[v(1+4v^2t_1t_2)],
\nonumber \\
&&h(k_1,q)=[v(t_1+t_2)k_1^2+v(4t_1+t_2)q^2+2v(2t_1-t_2)qk_1
\nonumber \\
&&+4i(1+v^2t_1t_2)q_\mu f_{\mu\nu}k_{1\nu}]/[4v(1+4v^2t_1t_2) ,
\nonumber \\
&&{\cal M}_2=\frac{1}{2v(1+4v^2t_1t_2)}\left(
\begin{array}{ccc}
 A_1 & -M_2 & M_2 \\
-M^T_2 & A_2 & -N_2 \\
M^T_2 & -N^{\rm T}_2 & A_3
\end{array} \right),
\nonumber \\
&& A_1^{\mu\nu}=[2v(t_1+t_2)+s_1(1+4v^2t_1t_2)]\delta_{\mu\nu},
\nonumber \\
&& A_2^{\mu\nu}=[2v(t_1+t_2)+s_2(1+4v^2t_1t_2)]\delta_{\mu\nu},
\nonumber \\
&& A_3^{\mu\nu}=[2v(t_1+t_2)+s_3(1+4v^2t_1t_2)]\delta_{\mu\nu},
\nonumber \\
&&M^{\mu\nu}_2= 2v(t_1-t_2)\delta^{\mu\nu}-i(1-4v^2t_1t_2)f^{\mu\nu},
\nonumber \\
&&N_2^{\mu\nu}= 2v(t_1+t_2)\delta^{\mu\nu}-i(1+4v^2t_1t_2)f^{\mu\nu},
\nonumber
\end{eqnarray}
and ${\cal W}_{\mu}$ has the structure:
\begin{eqnarray}
&&{\cal W}_\mu =B_{\mu,\nu\alpha\beta}
p^{\nu}_1  p^{\alpha}_2  p^{\beta}_3
-\sum_{i=1}^3 C^i_{\mu\nu}p^\nu_i, \nonumber \\
&&C^1_{\mu\nu}=\frac{m_u^2}{(1-s^2_2)(1-s^2_3)}
(1+s_1s_2+s_1s_3+s_2s_3)
[\delta_{\mu\nu}(1-s_2s_3)+if_{\mu\nu}(s_3-s_2)],
\nonumber \\
&&C^2_{\mu\nu}=\frac{m_u^2}{(1-s^2_1)(1-s^2_3)}
(1+s_1s_2+s_1s_3+s_2s_3)
[\delta_{\mu\nu}(1+s_1s_3)+if_{\mu\nu}(s_1+s_3)],
\nonumber \\
&&C^3_{\mu\nu}=\frac{m_u^2}{(1-s^2_1)(1-s^2_2)}
(1+s_1s_2+s_1s_3+s_2s_3)
[\delta_{\mu\nu}(1+s_1s_2)+if_{\mu\nu}(s_1+s_2)],
\nonumber \\
&&B_{\mu,\nu\alpha\beta}=
(\delta_{\alpha\beta}-is_1f_{\alpha\beta})
[\delta_{\mu\nu}(s_2s_3-1)+if_{\mu\nu}(s_2-s_3)] \nonumber \\
&&+(\delta_{\nu\beta}+is_2f_{\nu\beta})
[\delta_{\mu\alpha}(1+s_1s_3)+if_{\mu\alpha}(s_1+s_3)] \nonumber \\
&&+(\delta_{\alpha\nu}+is_3f_{\alpha\nu})
[\delta_{\beta\mu}(1+s_1s_2)+if_{\beta\mu}(s_1+s_2)]. \nonumber
\end{eqnarray}
Integrating over momenta $p_1$, $p_2$ and $p_3$,
and, then,  averaging over different configurations of the vacuum field
by means of formulas ($k_1^2=-M_\pi^2$, $k_2^2=-M_\pi^2$)
\begin{eqnarray}
\label{avec}
&&\int d\sigma_{B}\exp(2i\psi k_{1\gamma}f_{\gamma\sigma}q_\sigma)=
{\cal F}_0(\psi;q^2,M_\pi^2),
\\
&&\int d\sigma_{B} if_{\mu}(q)\exp(2i\psi k_{1\gamma}f_{\gamma\sigma}q_\sigma)
=(k_1+k_2)_\mu q^2{\cal F}_1(\psi;q^2,M_\pi^2),
\nonumber\\
&&\int d\sigma_{B} k_{1\alpha} q_\nu q_\beta
f_{\mu\nu}f_{\alpha\beta}\exp(2i\psi k_{1\gamma}f_{\gamma\sigma}q_\sigma)=
-(k_1+k_2)_\mu q^2{\cal F}_2(\psi;q^2,M_\pi^2)/2,
\nonumber\\
&&{\cal F}_0=\frac{1}{\psi\sqrt{q^2(q^2+4M^2_\pi)}}
\sinh\left[\psi\sqrt{q^2(q^2+4M^2_\pi)}\right],
\nonumber\\
&&{\cal F}_1=\frac{1}{\psi q^2(q^2+4M^2_\pi)}
\left[\cosh\left[\psi\sqrt{q^2(q^2+4M^2_\pi)}\right]
-{\cal F}_0\right],
\nonumber\\
&&{\cal F}_2={\cal F}_0+4{\cal F}_1/\psi,
\nonumber\\
\label{psi}
&&\psi=[s_1s_2s_3+vt_1s_2(s_1+s_3)+vt_2s_3(s_1+s_2)
+v^2t_1t_2(s_1+s_2+s_3+s_1s_2s_3)]/2v\chi,
\nonumber\\
&&\chi=2v(t_1+t_2)(1+s_1s_2+s_1s_3+s_2s_3)
+(1+4v^2t_1t_2)(s_1+s_2+s_3+s_1s_2s_3),
\nonumber
\end{eqnarray}
we get Eq.~(\ref{3}) for
$F^\triangle_\pi$.
Functions
$\Phi_i(M^2_\pi,q^2;s,t)$ (i=1,2,3) in Eq.~(\ref{3})
have the form
\begin{eqnarray}
&&\Phi_i=[{\cal F}_0P_i(s,t)
+{\cal F}_1R_i(M^2_\pi,q^2;s,t)
+{\cal F}_2L_i(q^2;s,t))]/8v\chi^5.
\nonumber
\end{eqnarray}
\begin{eqnarray}
&&R_j=q^2\sum_{k=1,2,3}Q_{jk}\chi_k, \ \ \ (j=1,3),
\nonumber \\
&&R_2=q^2\left[\frac{Q_{21}}{(1-s_1^2)(1-s_2^2)}
+\frac{Q_{22}}{(1-s_1^2)(1-s_3^2)}
+\frac{Q_{23}}{(1-s_2^2)(1-s_3^2)}\right],
\nonumber \\
&&R_4=q^2(q^2+4M_\pi^2)\sum_{j=1,2,3}Q_{4j}\psi_j+q^2 Q_{44},
\nonumber \\
&&P_j=\sum_{k=1,2,3}Q_{jk}\psi_k, \ \  \ (j=1,3)
\nonumber \\
&&P_2=\frac{Q_{24}}{(1-s_1^2)(1-s_2^2)}
+\frac{Q_{25}}{(1-s_1^2)(1-s_3^2)}
+\frac{Q_{26}}{(1-s_2^2)(1-s_3^2)},
\nonumber \\
&&P_4=Q_{45}, \
L_1=L_2=L_3=0, \  L_4=-q^2\sum_{k=1,2,3}Q_{4k}\chi_k.
\nonumber
\end{eqnarray}
\begin{eqnarray}
&&Q_{11}=(1+s_1s_2+s_1s_3+s_2s_3)
[1+4v^2t_1t_2+2v(t_1+t_2)s_1+2v^2t_1t_2(1+4v^2t_1t_2)(1-s_1^2)]
\nonumber\\
&&+(1+4v^2t_1t_2)(1-s_1^2)[2vt_1s_2(1+s_1s_3)+2vt_2s_3(1+s_1s_2)+s_2s_3-1]
\nonumber\\
&&+v(t_1+t_2)(1-s_1^2)\chi,
\nonumber\\
&&Q_{12}=-(1+s_1s_2+s_1s_3+s_2s_3)
\nonumber\\
&&\times\Bigl[2(1-s_1^2)v^2t_2^2(1+4v^2t_1^2)
+(1+4v^2t_1t_2+2v(t_1+t_2)s_2)(1+2vs_2t_1)\Bigr]
\nonumber\\
&&+(1-s_2^2)\Bigl[1+s_2s_3+s_3(s_1+s_2)+
4vt_1s_2(1+s_1s_3)+vt_2(s_3(1+s_1s_2)-(s_1+s_2))\nonumber\\
&&-4v^2t_1t_2\left[s_2s_3-1+s_3(s_2-s_1)+vt_2(2s_3(1+s_1s_2)+
s_2(1+s_1s_3)+s_1+s_3)\right]\Bigr],
\nonumber \\
&&Q_{21}=(1+4v^2t_1t_2)(1-s_1^2)(1-s_2^2)
+v(t_1-t_2)(s_2(1+s_1s_2)+s_1) \nonumber\\
&&-(1+v(t_1+t_2)s_1)(s_2(s_1+s_2)+1),
\nonumber\\
&&Q_{24}=-2\chi^2[-v(t_1-t_2)s_1(s_2(1+s_1s_2)+s_1)
+v(t_1+t_2)(s_2(s_1+s_2)+1)
\nonumber\\
&&+s_1(s_2(s_1+s_2)+1+s_1s_2)],
\nonumber\\
&&Q_{23}=(1+4v^2t_1t_2)(1-s_2^2)(1-s_3^2)
-v(t_1-t_2)(s_3-s_2)(s_2s_3-1)
\nonumber\\
&&-v(t_1+t_2)(s_3+s_2)(s_2s_3-1)
+2s_2^2(1-s_3^2)+2s_3^2(1-s_2^2)-(1-s_2s_3)(1+s_2s_3),
\nonumber\\
&&Q_{26}=-2\chi^2[v(t_1-t_2)(s_2-s_3)(s_2+s_3)
+v(t_1+t_2)(1-s_2s_3)(1+s_2s_3)
\nonumber \\
&&+(s_2+s_3)(1-s_2s_3)],
\nonumber \\
&&Q_{31}=2(1+s_1s_2+s_1s_3+s_2s_3)
[vt_1(1-s_1)+vt_2(1+s_1)+s_1]
\nonumber\\
&&[vt_1(1+s_1)+vt_2(1-s_1)+s_1],
\nonumber\\
&&Q_{32}=-2(1-s_2^2)\biggl[(s_1+v(t_1+t_2))
[s_2+s_3+v(t_1+t_2)(1+s_2s_3)]
\nonumber \\
&&+v^2s_1(t_1-t_2)^2(s_2+s_3)\biggr],
\nonumber \\
&&Q_{41}=(1+s_1s_2+s_1s_3+s_2s_3)
[s_1(1+4v^2t_1t_2)+v(t_1+t_2)(1+s_1^2)]
\nonumber\\
&&-(1+4v^2t_1t_2)(1-s_1^2)\Bigl[(1-s_2s_3)(v(t_1+t_2)+s_1)
+vs_1(t_1-t_2)(s_3-s_2)\Bigr],
\nonumber\\
&&Q_{42}=(1+s_1s_2+s_1s_3+s_2s_3)
(s_2+2vt_1)[1+4v^2t_1t_2+2v(t_1+t_2)s_2]
\nonumber\\
&&+(1-s_2^2)\Bigl[s_2(s_1s_3-1)
+2vt_1[s_1s_2-1+s_3(s_2-s_1)]+2vt_2s_3(s_1+s_2)\nonumber\\
&&-4v^2t_1t_2[s_1(s_2s_3-1)+s_3(s_1s_2-1)+vt_1(1+s_1s_2+s_1s_3+s_2s_3)
\nonumber\\
&&+vt_2(s_3(s_1+s_2)-(1+s_1s_2))]\Bigr],
\nonumber \\
&&Q_{44}=4v\chi[d_{41}+vt_1 d_{42}+vt_2 d_{43}],
\nonumber\\
&&Q_{45}=-4v\chi[d_{44}+vt_1 d_{45}+vt_2 d_{46}],
\nonumber\\
&&d_{41}=4v(t_1-t_2)(s_2-s_3)(1+s_1s_2+s_1s_3+s_2s_3)
\nonumber\\
&&+(1-4v^2t_1t_2)\Bigl[2(1+s_1s_2+s_1s_3+s_2s_3)
\Bigl((1+4v^2t_1t_2)(1+s_1s_2+s_1s_3+s_2s_3)
\nonumber\\
&&-2(1-s_2s_3)\Bigr)+(1+4v^2t_1t_2)(1-s_1^2)(1-s_2^2)(1-s_3^2)\Bigr]
\nonumber\\
&&+(1+4v^2t_1t_2)(1-s_1^2)(1-s_2s_3)(1+s_2s_3),
\nonumber\\
&&d_{42}=4(1+s_1s_2+s_1s_3+s_2s_3)
\Bigl[v(t_1-t_2)[s_1(s_2-s_3)+s_2s_3-1]
\nonumber\\
&&+2(1-4v^2t_1t_2)s_2(1+s_1s_3)\Bigr]
+2vt_1(1+4v^2t_1t_2)s_2(1-s_1^2)(1-s_3^2),
\nonumber \\
&&d_{44}=4s_1(1+s_1s_2+s_1s_3+s_2s_3)
[v(t_1-t_2)(s_3-s_2)+(1-4v^2t_1t_2)(1-s_2s_3)]
\nonumber\\
&&+(1+4v^2t_1t_2)(1-s_1^2)(s_2+s_3)(1-s_2s_3)
\nonumber\\
&&d_{45}=4(1+s_1s_2+s_1s_3+s_2s_3)
\Bigl[v(t_1-t_2)[s_1(1-s_2s_3)+s_3-s_2]
\nonumber\\
&&+(1-4v^2t_1t_2)[s_1(s_3-s_2)+1-s_2s_3]\Bigr]
+(1+4v^2t_1t_2)(1-s_1^2)(1+s_2^2)(1-s_3^2),
\nonumber \\
&&\chi_1=-4v^2t_1t_2(1-s_2^2)(1-s_3^2)-s_2(1-s_3^2)(s_2+2vt_1)
-s_3(1-s_2^2)(s_3+2vt_2),
\nonumber\\
&&\chi_2=-(1+4v^2t_1t_2)(1-s_1^2)(1-s_3^2)
+2vt_1(s_1+s_3)(1+s_1s_3)-2vt_2s_3(1-s_1^2)
\nonumber\\
&&+s_3(s_1+s_3)+s_1s_3+1,
\nonumber\\
&&\psi_1=(1-s_2s_3)[s_2+s_3+v(t_1+t_2)(1+s_2s_3)]
+v(t_1-t_2)(s_2-s_3)(s_2+s_3),
\nonumber\\
&&\psi_2=[1+s_1s_3+s_3(s_1+s_3)](s_1+v(t_1+t_2))
+v(t_1-t_2)[s_1+s_3+s_3(1+s_1s_3)],
\nonumber
\end{eqnarray}
The rest of functions can be obtained by the permutations
of arguments
$t_1 \leftrightarrow t_2, s_2 \leftrightarrow s_3$:
\begin{eqnarray}
&&Q_{13}(t_1,t_2,s_1,s_2,s_3)=Q_{12}(t_2,t_1,s_1,s_3,s_2), \
Q_{22}(t_1,t_2,s_1,s_2,s_3)=Q_{21}(t_2,t_1,s_1,s_3,s_2),
\nonumber\\
&&Q_{25}(t_1,t_2,s_1,s_2,s_3)=Q_{24}(t_2,t_1,s_1,s_3,s_2), \
Q_{33}(t_1,t_2,s_1,s_2,s_3)=Q_{32}(t_2,t_1,s_1,s_3,s_2),
\nonumber\\
&&Q_{43}(t_1,t_2,s_1,s_2,s_3)=Q_{42}(t_2,t_1,s_1,s_3,s_2), \
\nonumber\\
&&d_{43}(t_1,t_2,s_1,s_2,s_3)=d_{42}(t_2,t_1,s_1,s_3,s_2), \
d_{46}(t_1,t_2,s_1,s_2,s_3)=d_{45}(t_2,t_1,s_1,s_3,s_2),
\nonumber\\
&&\chi_3(t_1,t_2,s_1,s_2,s_3)= \chi_2(t_2,t_1,s_1,s_3,s_2),
\psi_3(t_1,t_2,s_1,s_2,s_3)=\psi_2(t_2,t_1,s_1,s_3,s_2).
\nonumber
\end{eqnarray}

In the configuration space the $\pi^+\pi^-\gamma$ vertex
corresponding to the bubble diagram in Fig.~5  looks as
\begin{eqnarray}
\label{LB}
&&\Lambda_\mu^\circ(x,y,z)=
h^2_\pi{\rm Tr}([\tau^+,\tau^-]_{-}{\cal Q})\int d\sigma_{B}
{\rm Tr}\{i\gamma_5 F_{00}(x)S(x,y)i\gamma_5
\Gamma_\mu(z,y)S(y,x)\}.
\nonumber
\end{eqnarray}
A calculation of this vertex is analogous to
above described case of $\Lambda_\mu^\triangle$.
Using the expression for $\Gamma_\mu(z,y)$~(\ref{finalb}) we
get Eq.~(\ref{b})  with
\begin{eqnarray}
&&\Phi_i^\circ=({\cal F}_0(\psi^\circ;q^2,M_\pi^2) P_i^\circ(s,t)
+{\cal F}_1(\psi^\circ;q^2,M_\pi^2) R_i^\circ(q^2;s,t))/(\chi^\circ)^4.
\nonumber
\end{eqnarray}
$P_i^\circ,R_i^\circ$ are the polynomials of $t$, $s_1$, $s_2$ and $s_3$,
\begin{eqnarray}
&&R_1^\circ=q^2\chi^\circ (1+s_1s_2)^2(s_1s_2+vt_1(s_1+s_2))t_2/2,
\nonumber \\
&&R_2^\circ=q^2[3v(1-4v^2t_1t_2)(s_1s_2+vt_1(s_1+s_2))-v^2t_1\chi^\circ]
t_2(1+s_1s_2),
\nonumber \\
&&P_1^\circ=\chi^\circ (1+s_1s_2)^2(s_1-s_2)t_2/2
\nonumber \\
&&P_2^\circ=3v(1-4v^2t_1t_2)(s_1-s_2)t_2(1+s_1s_2)/2,
\nonumber
\end{eqnarray}
and
\begin{eqnarray}
&&{\cal F}_0=\frac{1}{\psi^\circ\sqrt{q^2(q^2+4M^2_\pi)}}
\sinh\left[\psi^\circ\sqrt{q^2(q^2+4M^2_\pi)}\right],
\nonumber \\
&&{\cal F}_1=\frac{1}{\psi^\circ q^2(q^2+4M^2_\pi)}
\left\{\cosh\left[\psi^\circ\sqrt{q^2(q^2+4M^2_\pi)}\right]
-{\cal F}_0\right\},
\nonumber \\
&&\psi^\circ=vt_2\beta(s_1s_2+vt_1(s_1+s_2))/2v\chi^\circ,
\nonumber \\
&&\chi^\circ=2v(t_1+t_2)(1+s_1s_2)+(1+4v^2t_1t_2)(s_1+s_2).
\nonumber
\end{eqnarray}

\newpage

\begin{table}
\caption{Parameters of the model $(\Lambda^4=3B^2)$.}
\label{par}
\vspace*{3mm}

\begin{tabular}{|ccccccc|}
\hline
$m_u$ (MeV)
&$m_d$ (MeV)&$m_s$ (MeV)&$m_c$ (MeV)&$m_b$ (MeV)&$\Lambda$ (MeV)&$g$  \\
\hline
198.3&198.3&413&1650&4840&319.5&9.96\\
\hline
\end{tabular}
\end{table}

\newpage

\begin{table}
\begin{center}
\begin{minipage}{120.mm}
\caption{The two-photon decay constant $g_{\pi\gamma\gamma}$
(${\rm Gev}^{-1}$) and decay width $\Gamma(\pi^0\to\gamma\gamma)$
($\rm ev$); $g_{\pi\gamma\gamma}^\ast, \Gamma^\ast$ are the values
calculated without taking into account the spin-field interaction}
\end{minipage}
\vspace*{1cm}

\begin{tabular}{|c|c|c||c|c|c|}\hline
$g_{\pi\gamma\gamma}$&
$g_{\pi\gamma\gamma}^\ast$ &$g_{\pi\gamma\gamma}^{\rm exp}$ \cite{bebek}&
$\Gamma$&$\Gamma^\ast$&$\Gamma^{\rm exp}$ \cite{bebek}
\\ \hline
0.235&0.108&0.276&6.3&1.34&8.74
\\ \hline
\end{tabular}
\end{center}
\end{table}

\newpage

\begin{figure}[h]
\centering
~\\[2cm]
\vspace*{1cm}

\unitlength=1.00mm
\special{em:linewidth 0.4pt}
\linethickness{0.4pt}
\begin{picture}(105.00,150.00)
\put(20.00,150.00){\line(0,-1){80.00}}
\put(20.00,70.00){\line(1,0){80.00}}
\put(100.00,70.00){\line(1,0){5.00}}
\put(105.00,70.00){\line(-1,0){5.00}}
\put(100.00,70.00){\line(0,1){10.00}}
\put(100.00,80.00){\line(-1,0){25.00}}
\put(75.00,80.00){\line(0,1){20.00}}
\put(75.00,100.00){\line(-1,0){19.00}}
\put(56.00,100.00){\line(0,1){45.00}}
\put(56.00,145.00){\line(-1,0){36.00}}
\put(20.00,145.00){\line(0,1){0.00}}
\put(20.00,145.00){\line(0,0){0.00}}
\put(56.00,100.00){\line(0,-1){3.00}}
\put(56.00,94.00){\line(0,-1){4.00}}
\put(56.00,86.00){\line(0,-1){5.00}}
\put(56.00,78.00){\line(0,-1){4.00}}
\put(56.00,72.00){\line(0,-1){2.00}}
\put(75.00,80.00){\line(0,-1){3.00}}
\put(75.00,73.00){\line(0,-1){3.00}}
\put(26.00,140.00){\makebox(0,0)[cc]{$\pi$}}
\put(34.00,140.00){\makebox(0,0)[cc]{$\rho$}}
\put(41.00,140.00){\makebox(0,0)[cc]{$K$}}
\put(49.00,140.00){\makebox(0,0)[cc]{$K^*$}}
\put(25.00,128.00){\makebox(0,0)[cc]{$\omega$}}
\put(33.00,128.00){\makebox(0,0)[cc]{$\phi$}}
\put(41.00,128.00){\makebox(0,0)[cc]{$b_1$}}
\put(49.00,128.00){\makebox(0,0)[cc]{$K_1$}}
\put(25.00,116.00){\makebox(0,0)[cc]{$\eta_c$}}
\put(33.00,116.00){\makebox(0,0)[cc]{$\psi$}}
\put(41.00,116.00){\makebox(0,0)[cc]{$\psi^\prime$}}
\put(49.00,116.00){\makebox(0,0)[cc]{$\psi^{\prime\prime}$}}
\put(25.00,103.00){\makebox(0,0)[cc]{$\chi_{c0}$}}
\put(33.00,103.00){\makebox(0,0)[cc]{$\chi_{c1}$}}
\put(42.00,103.00){\makebox(0,0)[cc]{$\chi_{c2}$}}
\put(49.00,103.00){\makebox(0,0)[cc]{$\Upsilon$}}
\put(25.00,91.00){\makebox(0,0)[cc]{$\Upsilon^\prime$}}
\put(33.00,91.00){\makebox(0,0)[cc]{$\Upsilon^{\prime\prime}$}}
\put(49.00,91.00){\makebox(0,0)[cc]{$\chi_{b1}$}}
\put(41.00,91.00){\makebox(0,0)[cc]{$\chi_{b0}$}}
\put(25.00,83.00){\makebox(0,0)[cc]{$\chi_{b2}$}}
\put(33.00,83.00){\makebox(0,0)[cc]{$\chi_{b0}^\prime$}}
\put(41.00,83.00){\makebox(0,0)[cc]{$\chi_{b1}^\prime$}}
\put(49.00,83.00){\makebox(0,0)[cc]{$\chi_{b2}^\prime$}}
\put(29.00,75.00){\makebox(0,0)[cc]{$D^*$}}
\put(37.00,75.00){\makebox(0,0)[cc]{$D_s^*$}}
\put(45.00,75.00){\makebox(0,0)[cc]{$B_s^*$}}
\put(61.00,96.00){\makebox(0,0)[cc]{$a_1$}}
\put(69.00,96.00){\makebox(0,0)[cc]{$a_2$}}
\put(61.00,90.00){\makebox(0,0)[cc]{$K_1^*$}}
\put(69.00,90.00){\makebox(0,0)[cc]{$K_2^*$}}
\put(61.00,83.00){\makebox(0,0)[cc]{$D$}}
\put(69.00,83.00){\makebox(0,0)[cc]{$D_s$}}
\put(65.00,76.00){\makebox(0,0)[cc]{$B^*$}}
\put(81.00,75.00){\makebox(0,0)[cc]{$B$}}
\put(91.00,75.00){\makebox(0,0)[cc]{$B_s$}}
\put(100.00,68.00){\makebox(0,0)[cc]{$8\%$}}
\put(75.00,68.00){\makebox(0,0)[cc]{$5\%$}}
\put(56.00,68.00){\makebox(0,0)[cc]{$3\%$}}
\put(17.00,80.00){\makebox(0,0)[cc]{$2$}}
\put(17.00,100.00){\makebox(0,0)[cc]{$7$}}
\put(16.00,145.00){\makebox(0,0)[cc]{$27$}}
\put(24.00,150.00){\makebox(0,0)[cc]{$N$}}
\put(50.00,60.00){\makebox(0,0)[lc]{$\frac{|M_{\rm exp}-M|}{M_{\rm exp}}100\%$}}
\end{picture}

\vspace*{-2cm}
\caption{
 A diagram illustrating accuracy of description of meson spectrum.
        All the masses were calculated with the values of parameters
        given in Table~\ref{par}. Parameters were fitted using the masses
        of $\pi$, $\rho$, $K$, $K^*$, $\psi$ and $\Upsilon$ mesons.}

     \end{figure}
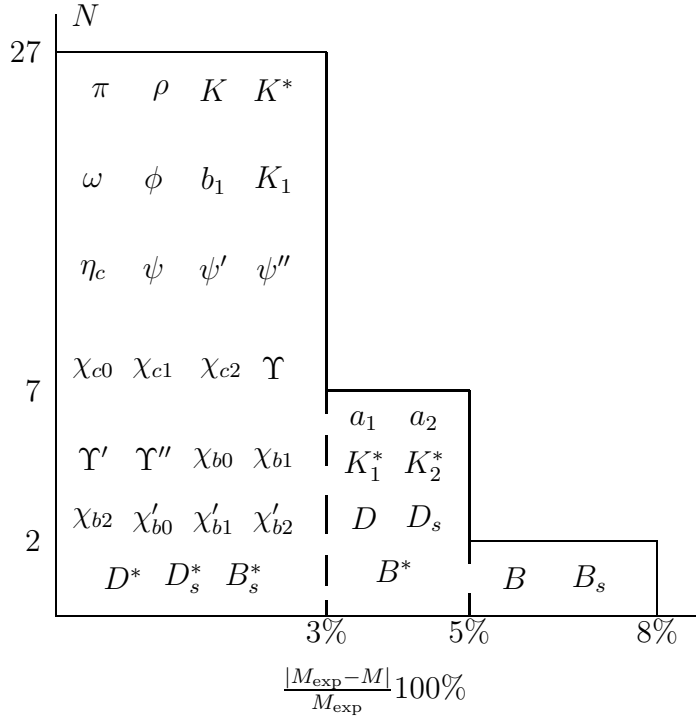
\newpage

\begin{figure}[h]
\centering
~\\[-2cm]
\unitlength=1.00mm
\special{em:linewidth 0.4pt}
\linethickness{0.4pt}
\begin{picture}(152.00,150.00)
\put(20.00,121.00){\circle*{4.47}}
\put(67.00,121.00){\circle{4.47}}
\put(91.00,121.00){\circle{4.47}}
\put(124.00,121.00){\circle{4.47}}
\put(9.00,121.00){\line(1,0){10.00}}
\put(19.00,121.00){\line(0,-1){1.00}}
\put(19.00,120.00){\line(-1,0){10.00}}
\put(20.00,120.00){\line(2,-5){3.33}}
\put(21.00,120.00){\line(2,-1){6.00}}
\put(67.00,119.00){\line(1,-2){3.00}}
\put(69.00,120.00){\line(5,-3){5.00}}
\put(91.00,119.00){\line(1,-2){3.00}}
\put(93.00,120.00){\line(5,-3){5.00}}
\put(124.00,119.00){\line(5,-6){5.00}}
\put(126.00,120.00){\line(2,-1){6.00}}
\put(67.00,123.00){\line(0,1){2.00}}
\put(67.00,126.00){\line(0,1){2.00}}
\put(67.00,129.00){\line(0,1){2.00}}
\put(60.00,121.00){\line(1,0){5.00}}
\put(65.00,121.00){\line(0,-1){1.00}}
\put(65.00,120.00){\line(-1,0){5.00}}
\put(83.00,121.00){\line(1,0){6.00}}
\put(89.00,121.00){\line(0,-1){1.00}}
\put(89.00,120.00){\line(-1,0){6.00}}
\put(116.00,121.00){\line(1,0){6.00}}
\put(122.00,121.00){\line(0,-1){1.00}}
\put(122.00,120.00){\line(-1,0){6.00}}
\put(28.00,121.00){\makebox(0,0)[cc]{$=$}}
\put(78.00,121.00){\makebox(0,0)[cc]{$+$}}
\put(99.00,121.00){\makebox(0,0)[cc]{$+$}}
\put(112.00,121.00){\makebox(0,0)[cc]{$+$}}
\put(102.00,120.00){\circle*{0.00}}
\put(105.00,120.00){\circle*{0.00}}
\put(108.00,120.00){\circle*{0.00}}
\put(138.00,121.00){\makebox(0,0)[cc]{$+$}}
\put(144.00,120.00){\circle*{0.00}}
\put(147.00,120.00){\circle*{0.00}}
\put(150.00,120.00){\circle*{0.00}}
\put(9.00,140.00){\rule{14.00\unitlength}{1.00\unitlength}}
\put(29.00,141.00){\makebox(0,0)[cc]{$=$}}
\put(36.00,140.00){\line(1,0){14.00}}
\put(63.00,140.00){\line(1,0){17.00}}
\put(91.00,140.00){\line(1,0){15.00}}
\put(124.00,140.00){\line(1,0){19.00}}
\put(43.00,121.00){\circle{4.47}}
\put(43.00,119.00){\line(3,-5){4.33}}
\put(45.00,119.00){\line(5,-2){5.00}}
\put(37.00,121.00){\line(1,0){4.00}}
\put(41.00,121.00){\line(0,-1){1.00}}
\put(41.00,120.00){\line(-1,0){4.00}}
\put(53.00,121.00){\makebox(0,0)[cc]{$+$}}
\put(72.00,140.00){\line(0,1){2.00}}
\put(72.00,144.00){\line(0,1){3.00}}
\put(96.00,140.00){\line(0,1){2.00}}
\put(96.00,143.00){\line(0,1){2.00}}
\put(96.00,146.00){\line(0,1){3.00}}
\put(101.00,140.00){\line(0,1){2.00}}
\put(101.00,144.00){\line(0,0){0.00}}
\put(101.00,144.00){\line(0,1){2.00}}
\put(101.00,147.00){\line(0,1){2.00}}
\put(127.00,140.00){\line(0,1){2.00}}
\put(127.00,143.00){\line(0,1){3.00}}
\put(127.00,147.00){\line(0,1){2.00}}
\put(130.00,140.00){\line(0,1){2.00}}
\put(130.00,143.00){\line(0,1){3.00}}
\put(130.00,147.00){\line(0,1){2.00}}
\put(139.00,140.00){\line(0,1){2.00}}
\put(139.00,144.00){\rule{0.00\unitlength}{2.00\unitlength}}
\put(139.00,148.00){\line(0,1){2.00}}
\put(132.00,145.00){\circle*{0.00}}
\put(134.00,145.00){\circle*{0.00}}
\put(136.00,145.00){\circle*{0.00}}
\put(57.00,140.00){\makebox(0,0)[cc]{$+$}}
\put(86.00,140.00){\makebox(0,0)[cc]{$+$}}
\put(110.00,140.00){\makebox(0,0)[cc]{$+$}}
\put(120.00,140.00){\makebox(0,0)[cc]{$+$}}
\put(113.00,140.00){\circle*{0.00}}
\put(115.00,140.00){\circle*{0.00}}
\put(117.00,140.00){\circle*{0.00}}
\put(148.00,140.00){\circle*{0.00}}
\put(150.00,140.00){\circle*{0.00}}
\put(152.00,140.00){\circle*{0.00}}
\put(152.00,140.00){\circle*{0.00}}
\put(146.00,140.00){\makebox(0,0)[cc]{$+$}}
\put(9.00,99.00){\rule{11.00\unitlength}{1.00\unitlength}}
\put(41.00,99.00){\line(1,0){11.00}}
\put(52.00,99.00){\line(0,0){0.00}}
\put(75.00,99.00){\line(1,0){4.00}}
\put(81.00,99.00){\line(1,0){3.00}}
\put(86.00,99.00){\line(1,0){4.00}}
\put(14.00,94.00){\makebox(0,0)[cc]{$S(x,y|A)$}}
\put(46.00,94.00){\makebox(0,0)[cc]{$S(x,y)$}}
\put(82.00,94.00){\makebox(0,0)[cc]{$A_\mu$}}
\put(13.00,112.00){\makebox(0,0)[cc]{$V_{\cal N}(x|A)$}}
\put(37.00,112.00){\makebox(0,0)[cc]{$V_{\cal N}$}}
\put(60.00,112.00){\makebox(0,0)[cc]{$V^{(1)}_{\cal N}$}}
\put(84.00,112.00){\makebox(0,0)[cc]{$V^{(2)}_{\cal N}$}}
\put(115.00,112.00){\makebox(0,0)[cc]{$V^{(n)}_{\cal N}$}}
\put(90.00,123.00){\line(0,1){3.00}}
\put(90.00,128.00){\line(0,1){2.00}}
\put(90.00,131.00){\line(0,1){2.00}}
\put(92.00,123.00){\line(0,1){3.00}}
\put(92.00,128.00){\line(0,1){2.00}}
\put(92.00,131.00){\line(0,1){2.00}}
\put(122.00,123.00){\line(0,1){3.00}}
\put(122.00,128.00){\line(0,1){3.00}}
\put(122.00,132.00){\line(0,1){3.00}}
\put(123.00,123.00){\line(0,1){3.00}}
\put(123.00,128.00){\line(0,1){3.00}}
\put(123.00,132.00){\line(0,1){3.00}}
\put(124.00,129.00){\circle*{0.00}}
\put(125.00,129.00){\circle*{0.00}}
\put(126.00,129.00){\circle*{0.00}}
\put(126.00,121.00){\line(1,2){1.00}}
\put(127.00,124.00){\line(0,1){2.00}}
\put(127.00,128.00){\line(0,1){3.00}}
\put(127.00,132.00){\line(0,1){3.00}}
\put(106.00,99.00){\line(1,0){15.00}}
\put(106.00,98.00){\line(1,0){15.00}}
\put(114.00,94.00){\makebox(0,0)[cc]{$\Phi_{\cal N}$}}
\end{picture}

\caption{
 A decomposition of
       propagators and vertices, Eqs.~(\ref{s-decomp},\ref{v-decomp})
      into a series in electromagnetic field $A$. A procedure for calculation
      of the vertices $V^{(n)}_{\cal N}$ is described in Appendix~A.}

     \end{figure}
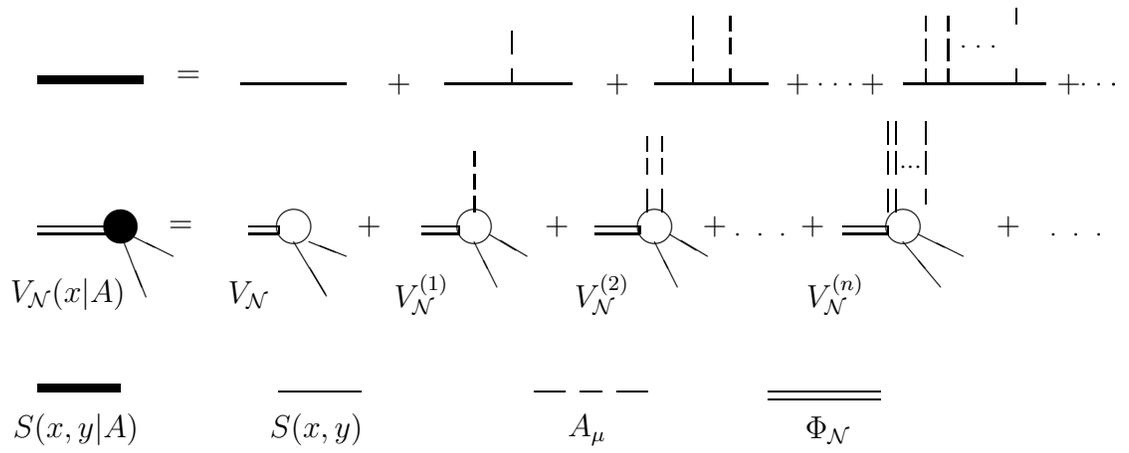

\newpage

\begin{figure}[h]
\centering
~\\[-2cm]
\unitlength=1mm
\special{em:linewidth 0.4pt}
\linethickness{0.4pt}
\begin{picture}(88.00,140.33)
\put(59.00,130.00){\circle{4.00}}
\put(61.00,131.00){\line(3,2){14.00}}
\put(75.00,140.33){\line(0,-1){18.33}}
\put(75.00,122.00){\line(-2,1){14.00}}
\put(75.00,140.00){\line(1,0){3.00}}
\put(80.00,140.00){\line(1,0){3.00}}
\put(85.00,140.00){\line(1,0){3.00}}
\put(75.00,122.00){\line(1,0){3.00}}
\put(80.00,122.00){\line(1,0){3.00}}
\put(85.00,122.00){\line(1,0){3.00}}
\put(46.00,131.00){\line(1,0){11.00}}
\put(46.00,129.00){\line(1,0){11.00}}
\put(51.00,126.00){\makebox(0,0)[cc]{$p$}}
\put(59.00,135.00){\makebox(0,0)[cc]{$t$}}
\put(66.00,123.00){\makebox(0,0)[cc]{$s_1$}}
\put(66.00,139.00){\makebox(0,0)[cc]{$s_3$}}
\put(79.00,131.00){\makebox(0,0)[cc]{$s_2$}}
\put(87.00,137.00){\makebox(0,0)[cc]{$k_1$}}
\put(87.00,118.00){\makebox(0,0)[cc]{$k_2$}}
\end{picture}

\caption{
The triangle diagram for the $\gamma^{\ast}\pi^0\to\gamma$
transition form factor.}

     \end{figure}
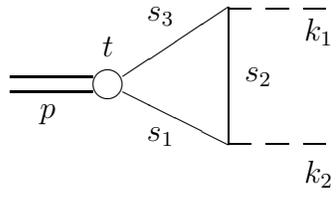

\newpage
\vspace*{10mm}

\begin{figure}[h]
\centering
~\\[-2cm]
\epsfig{file=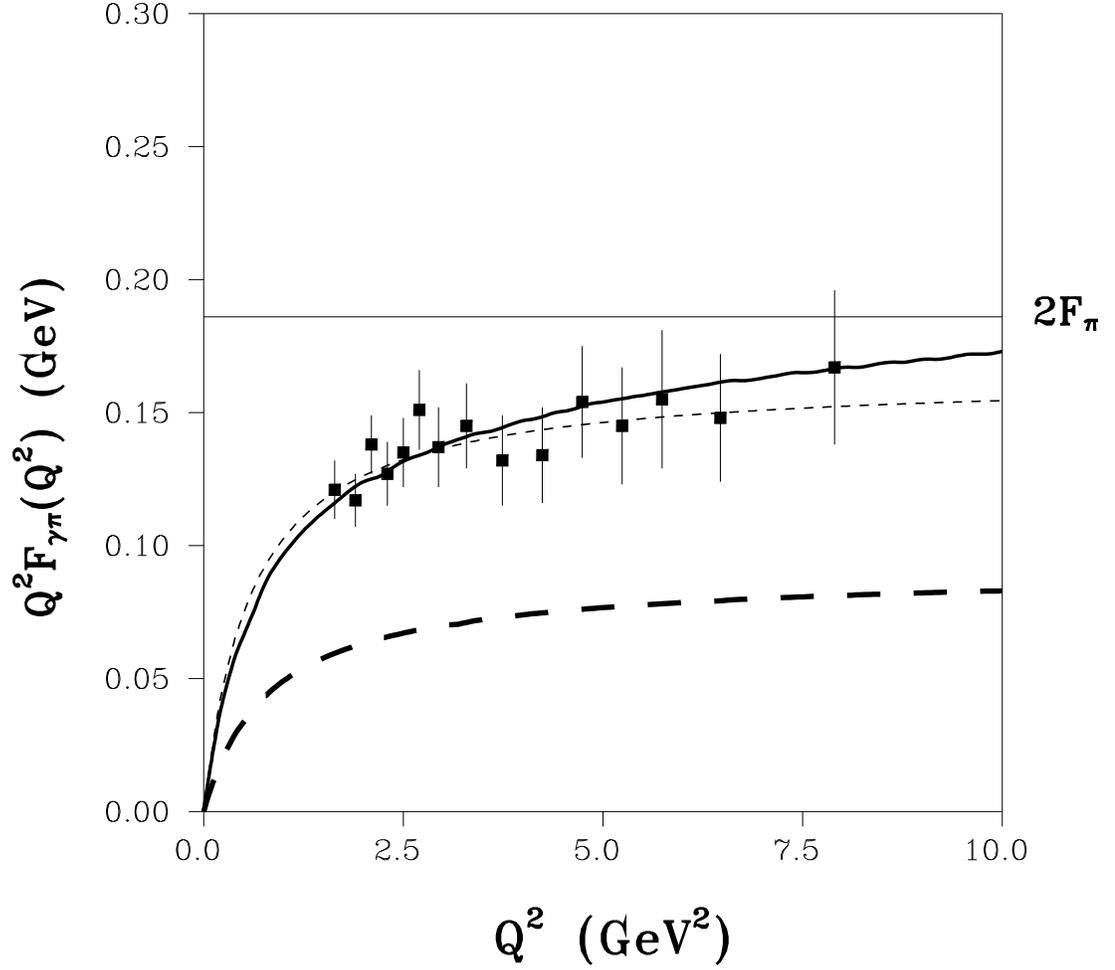,width=14.5cm}
~\\[1cm]
     \caption{Transition pion form factor.
      Solid curve represents a result of
      the model of induced nonlocal quark currents.
     Long dashed line - calculation without taking into account
     the spin-field interaction.
      Experimental fit \protect\cite{cleo} is given by short dashed line,
      and Brodsky-Lepage limit \protect\cite{br} ($\approx .186 \ {\rm Gev}$)
      is shown by solid straight line;
}
     \end{figure}

\newpage

\begin{figure}[h]
\centering
~\\[-2cm]
\unitlength=1.00mm
\special{em:linewidth 0.4pt}
\linethickness{0.4pt}
\begin{picture}(141.00,152.00)
\put(20.00,135.00){\circle{4.00}}
\put(41.00,135.00){\circle{4.00}}
\put(70.00,135.00){\circle{4.00}}
\put(90.00,135.00){\circle{4.00}}
\put(115.00,135.00){\circle{4.00}}
\put(134.00,135.00){\circle{4.00}}
\put(9.00,136.00){\line(1,0){9.00}}
\put(9.00,134.00){\line(1,0){9.00}}
\put(43.00,136.00){\line(1,0){8.00}}
\put(43.00,134.00){\line(1,0){8.00}}
\put(60.00,136.00){\line(1,0){8.00}}
\put(60.00,134.00){\line(1,0){8.00}}
\put(92.00,136.00){\line(1,0){9.00}}
\put(92.00,134.00){\line(1,0){9.00}}
\put(108.00,136.00){\line(1,0){5.00}}
\put(108.00,134.00){\line(1,0){5.00}}
\put(136.00,136.00){\line(1,0){5.00}}
\put(136.00,134.00){\line(1,0){5.00}}
\put(31.00,140.00){\line(0,1){2.00}}
\put(31.00,144.00){\line(0,1){3.00}}
\put(31.00,149.00){\line(0,1){3.00}}
\put(90.00,137.00){\line(0,1){3.00}}
\put(90.00,142.00){\line(0,1){3.00}}
\put(90.00,146.00){\line(0,1){3.00}}
\put(115.00,137.00){\line(0,1){3.00}}
\put(115.00,142.00){\line(0,1){3.00}}
\put(115.00,147.00){\line(0,1){3.00}}
\put(31.00,127.00){\makebox(0,0)[cc]{$s_1$}}
\put(40.00,143.00){\makebox(0,0)[cc]{$s_2$}}
\put(22.00,143.00){\makebox(0,0)[cc]{$s_3$}}
\put(34.00,151.00){\makebox(0,0)[cc]{$q$}}
\put(18.00,130.00){\makebox(0,0)[cc]{$t_1$}}
\put(43.00,130.00){\makebox(0,0)[cc]{$t_2$}}
\put(10.00,138.00){\makebox(0,0)[cc]{$k_1$}}
\put(49.00,138.00){\makebox(0,0)[cc]{$k_2$}}
\put(68.00,129.00){\makebox(0,0)[cc]{$t_1$}}
\put(81.00,127.00){\makebox(0,0)[cc]{$s_1$}}
\put(92.00,130.00){\makebox(0,0)[cc]{$t_2$}}
\put(60.00,139.00){\makebox(0,0)[cc]{$k_1$}}
\put(80.00,143.00){\makebox(0,0)[cc]{$s_2$}}
\put(100.00,138.00){\makebox(0,0)[cc]{$k_2$}}
\put(92.00,149.00){\makebox(0,0)[cc]{$q$}}
\put(92.00,139.00){\makebox(0,0)[cc]{$\beta$}}
\put(23.00,122.00){\makebox(0,0)[cc]{(a)}}
\put(101.00,122.00){\makebox(0,0)[cc]{(b)}}
\put(30.50,138.00){\oval(21.00,4.00)[t]}
\put(30.50,131.50){\oval(21.00,3.00)[b]}
\put(80.00,138.50){\oval(20.00,3.00)[t]}
\put(80.00,131.50){\oval(20.00,3.00)[b]}
\put(124.50,138.50){\oval(19.00,3.00)[t]}
\put(124.50,131.50){\oval(19.00,3.00)[b]}
\put(20.00,137.00){\line(0,1){1.00}}
\put(41.00,137.00){\line(0,1){1.00}}
\put(20.00,131.00){\line(0,1){2.00}}
\put(41.00,131.00){\line(0,1){2.00}}
\put(70.00,131.00){\line(0,1){1.00}}
\put(70.00,137.00){\line(0,0){0.00}}
\put(90.00,132.00){\line(0,1){1.00}}
\put(115.00,131.00){\line(0,1){2.00}}
\put(134.00,131.00){\line(0,1){2.00}}
\put(134.00,139.00){\line(0,-1){2.00}}
\put(70.00,131.00){\line(0,1){2.00}}
\put(70.00,138.00){\line(0,-1){1.00}}
\end{picture}

\caption{
The triangle (a) and bubble (b) diagrams
                     for pion charge form factor.}

     \end{figure}
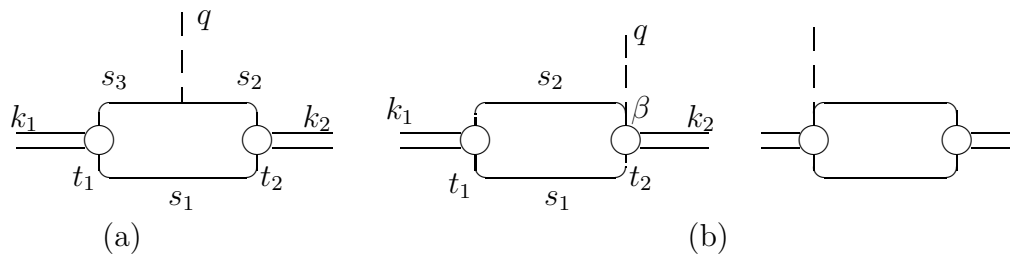



\begin{figure}[h]
\centering
~\\[-2cm]
\epsfig{file=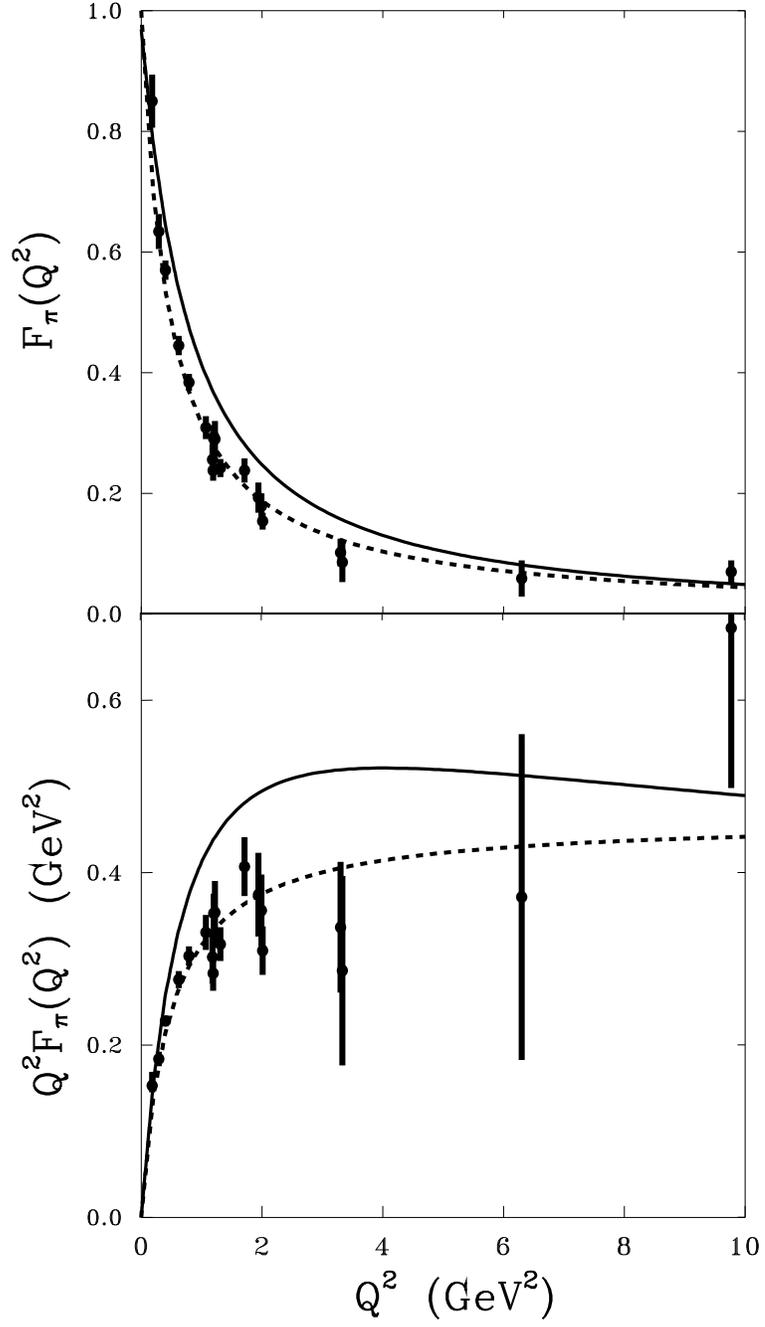,width=10cm}
~\\[1cm]
     \caption{The pion charge form factor $F_{\pi}(Q^2)$ calculated
     in the present model (solid line) compared with experimental fit
     (dashed line)}
     \end{figure}

\newpage

\begin{figure}[h]
\centering
~\\[-2cm]
\unitlength=1.00mm
\special{em:linewidth 0.4pt}
\linethickness{0.4pt}
\begin{picture}(91.00,154.00)
\put(66.00,125.00){\circle{4.00}}
\put(53.00,118.00){\circle{4.00}}
\put(78.00,118.00){\circle{4.00}}
\put(66.00,134.00){\circle{4.00}}
\put(65.00,127.00){\line(0,1){5.00}}
\put(67.00,132.00){\line(0,-1){5.00}}
\put(39.00,119.00){\line(1,0){12.00}}
\put(51.00,117.00){\line(-1,0){12.00}}
\put(80.00,119.00){\line(1,0){11.00}}
\put(80.00,117.00){\line(1,0){11.00}}
\put(66.00,144.00){\line(0,1){2.00}}
\put(66.00,148.00){\line(0,1){2.00}}
\put(66.00,152.00){\line(0,1){2.00}}
\put(70.00,149.00){\makebox(0,0)[cc]{$q$}}
\put(70.00,129.00){\makebox(0,0)[cc]{$\rho$}}
\put(45.00,122.00){\makebox(0,0)[cc]{$k_1$}}
\put(84.00,122.00){\makebox(0,0)[cc]{$k_2$}}
\put(65.50,113.50){\oval(25.00,5.00)[b]}
\put(58.50,122.50){\oval(11.00,5.00)[lt]}
\put(73.00,122.50){\oval(10.00,5.00)[rt]}
\put(62.50,139.00){\oval(3.00,10.00)[l]}
\put(69.50,139.00){\oval(3.00,10.00)[r]}
\put(59.00,125.00){\line(1,0){5.00}}
\put(68.00,125.00){\line(1,0){5.00}}
\put(63.00,144.00){\line(1,0){6.00}}
\put(53.00,122.00){\line(0,-1){2.00}}
\put(53.00,116.00){\line(0,-1){3.00}}
\put(78.00,114.00){\line(0,1){2.00}}
\put(78.00,122.00){\line(0,-1){2.00}}
\put(62.00,134.00){\line(1,0){2.00}}
\put(68.00,134.00){\line(1,0){2.00}}
\end{picture}

\caption{
The diagram with $\rho$-meson exchange.}
     \end{figure}
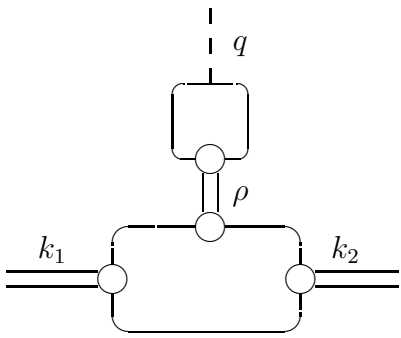

\end{document}